\documentclass[twocolumn]{aastex61}
\usepackage{silence}
\usepackage{graphicx}
\usepackage[caption=false]{subfig}

\usepackage{natbib}
\setcitestyle{authoryear,open={(},close={)}}

\usepackage{float}
\usepackage{color,soul}

\usepackage{wrapfig}
\usepackage{bm}
\usepackage{amsmath}

\usepackage{fancyhdr}
\usepackage{wasysym}

\newcommand{\fesc}{$f_{esc}^{Ly\alpha}$}
\newcommand{\OIII}{[O III] $\lambda5007$ }
\newcommand{\OII}{[O II] $\lambda3727$ }

\newcommand{\Orat}{[O III]/[O II] }
\newcommand{\fcov}{$f_{\mbox{\scriptsize cov}}$}
\newcommand{\mathcov}{f_{\mbox{\scriptsize cov}}}
\newcommand{\fcovLyA}{$f_{\mbox{\scriptsize cov}}^{\mbox{\scriptsize Ly}\alpha}$}
\newcommand{\mathcovLyA}{f_{\mbox{\scriptsize cov}}^{\mbox{\scriptsize Ly}\alpha}}
\newcommand{\NHI}{$N_{\mathrm{HI}}$}
\newcommand{\MHI}{$M_{\mathrm{HI}}$}
\newcommand{\FHI}{$f_{\mathrm{HI}}$}

\begin{document}

\title{Neutral Gas Properties and Ly$\alpha$ Escape in Extreme Green Pea Galaxies\footnote{Based on observations made with the NASA/ESA Hubble Space Telescope, obtained at the Space Telescope Science Institute, which is operated by the Association of Universities for Research in Astronomy, Inc., under NASA contract NAS 5-26555. These observations are associated with programs GO-14080.}}
\author{Jed H. McKinney} 
\affil{Department of Astronomy, University of Massachusetts, Amherst, MA 01003, USA.}
\email{jhmckinney@umass.edu}
\author{Anne E. Jaskot}
\altaffiliation{Hubble Fellow}
\affil{Department of Astronomy, University of Massachusetts, Amherst, MA 01003, USA.}
\author{M. S. Oey}
\affil{Department of Astronomy, University of Michigan, Ann Arbor, MI 48109, USA.}
\author{Min S. Yun}
\affil{Department of Astronomy, University of Massachusetts, Amherst, MA 01003, USA.}
\author{Tara Dowd}
\affil{Department of Astronomy, University of Massachusetts, Amherst, MA 01003, USA.}
\affil{The Chandra X-Ray Center, Cambridge, MA 02138, USA}
\author{James D. Lowenthal}
\affil{ Department of Astronomy, Smith College, Northampton, MA 01063, USA}

\begin{abstract}

Mechanisms regulating the escape of Ly$\alpha$ photons and ionizing radiation remain poorly understood. 
To study these processes we analyze VLA 21cm observations of one Green Pea (GP), J160810+352809 (hereafter J1608), and \textit{HST} COS spectra of 17 GP galaxies at $z<0.2$. All are highly ionized: J1608 has the highest [O III] $\lambda5007$/[O II] $\lambda3727$ for star-forming galaxies in SDSS, and the 17 GPs have \Orat $\geq6.6$. 
We set an upper limit on J1608's HI mass of $\log M_{HI}/M_\odot=8.4$, near or below average compared to similar mass dwarf galaxies.
In the COS sample, eight GPs show Ly$\alpha$ absorption components, six of which also have Ly$\alpha$ emission. 
The HI column densities derived from Ly$\alpha$ absorption are high, $\log N_{HI}/$cm$^{-2}=19-21$, well above the LyC optically thick limit. Using low-ionization absorption lines, we measure covering fractions (\fcov) of $0.1-1$, and find that \fcov\ strongly anti-correlates with Ly$\alpha$ escape fraction. Low covering fractions may facilitate Ly$\alpha$ and LyC escape through dense neutral regions. GPs with $\mathcov\sim1$ all have low neutral gas velocities, while GPs with lower $\mathcov=0.2-0.6$ have a larger range of velocities. Conventional mechanical feedback may help establish low \fcov\ in some cases, whereas other processes may be important for GPs with low velocities. 
Finally, we compare \fcov\ with proposed indicators of LyC escape. Ionizing photon escape likely depends on a combination of neutral gas geometry and kinematics, complicating the use of emission-line diagnostics for identifying LyC emitters.

\end{abstract}

\keywords{(cosmology:) dark ages, galaxies: dwarf, galaxies: evolution, galaxies: ISM, galaxies: starburst, galaxies: star clusters.}

\section{Introduction}

Determining the mechanisms responsible for reionizing the Universe at $z \sim 7-10$ remains an open question in observational cosmology. 
Active galactic nuclei (AGN) and massive stars can emit significant Lyman-continuum radiation (LyC, $\lambda<912$ \AA), but ionizing photon escape fractions from these sources into the intergalactic medium (IGM) are uncertain. 
High redshift quasar number counts indicate an established AGN population at the beginning of the epoch of re-ionization, but their ability to reproduce the LyC background is still debated (e.g. \citealt{Fontanot2012,F2014,Giallongo2015,MadauHaardt2015}). 
Cosmic reionization may instead be dominated by star-forming galaxies (SFGs) at $z>6$ (e.g. \citealt{Rob2015}). 
However, this scenario is complicated by large HI column densities around star-forming regions which prevent ionizing radiation escape into the IGM, despite high LyC photon fluxes from young, massive stars. Overcoming the neutral gas barrier may require either an ionized interstellar medium (ISM) or one perforated by optically thin channels. Both scenarios are plausible in low and intermediate mass galaxies (e.g. \citealt{JaskotOey2013,NakajimaOuchi2014,RivThorsen2015,Izotov2018b}). 

High-redshift observations of LyC-leaking SFGs are complicated by the effects of IGM attenuation and contamination by low-redshift galaxies (e.g. \citealt{Vanzella2012,Siana2015}). As a result, low-redshift studies of LyC emitters (LCEs) are necessary for inferring the physical mechanisms by which ionizing photons escape individual SFGs. 
Since their initial discovery by the Galaxy Zoo project from the Sloan Digital Sky Survey (SDSS) in 2009, a class of compact, luminous dwarfs known as Green Pea (GP; \citealt{Cardamone2009}) galaxies have become popular targets for their broad similarity to high-$z$ SFGs in terms of metallicity, line ratios and specific star formation rates (sSFRs), the ratio of SFR to stellar mass (e.g. \citealt{NakajimaOuchi2014,Izotov2011,Henry2015}). 

Green Peas have low stellar mass ($M_* < 10^{10}$ $M_\odot$), high sSFRs and redshifts $ z < 0.4$ \citep{Izotov2011, Izotov2017}. Originally selected by their characteristic green optical color, GPs have since been found to have very high \OIII / \OII ($O_{32}$) optical emission line ratios, suggestive of density-bounded nebulae (e.g. \citealt{JaskotOey2013}), and are generally metal-poor systems with mean oxygen abundances of $12+\log(\mbox{O/H})\sim 8.0$, roughly $1/5$ the solar value (e.g. \citealt{Izotov2011}). More GPs have since been discovered with the same $O_{32}$ and color properties as the original SDSS sample. A significant fraction show high Ly$\alpha$ escape fractions (\fesc) from $1-50\%$ (e.g. \citealt{Henry2015,Yang2017,Verhamme2017,Jaskot2017}).

 \begin{figure}[tb]
    \centering
    \includegraphics[width=\linewidth]{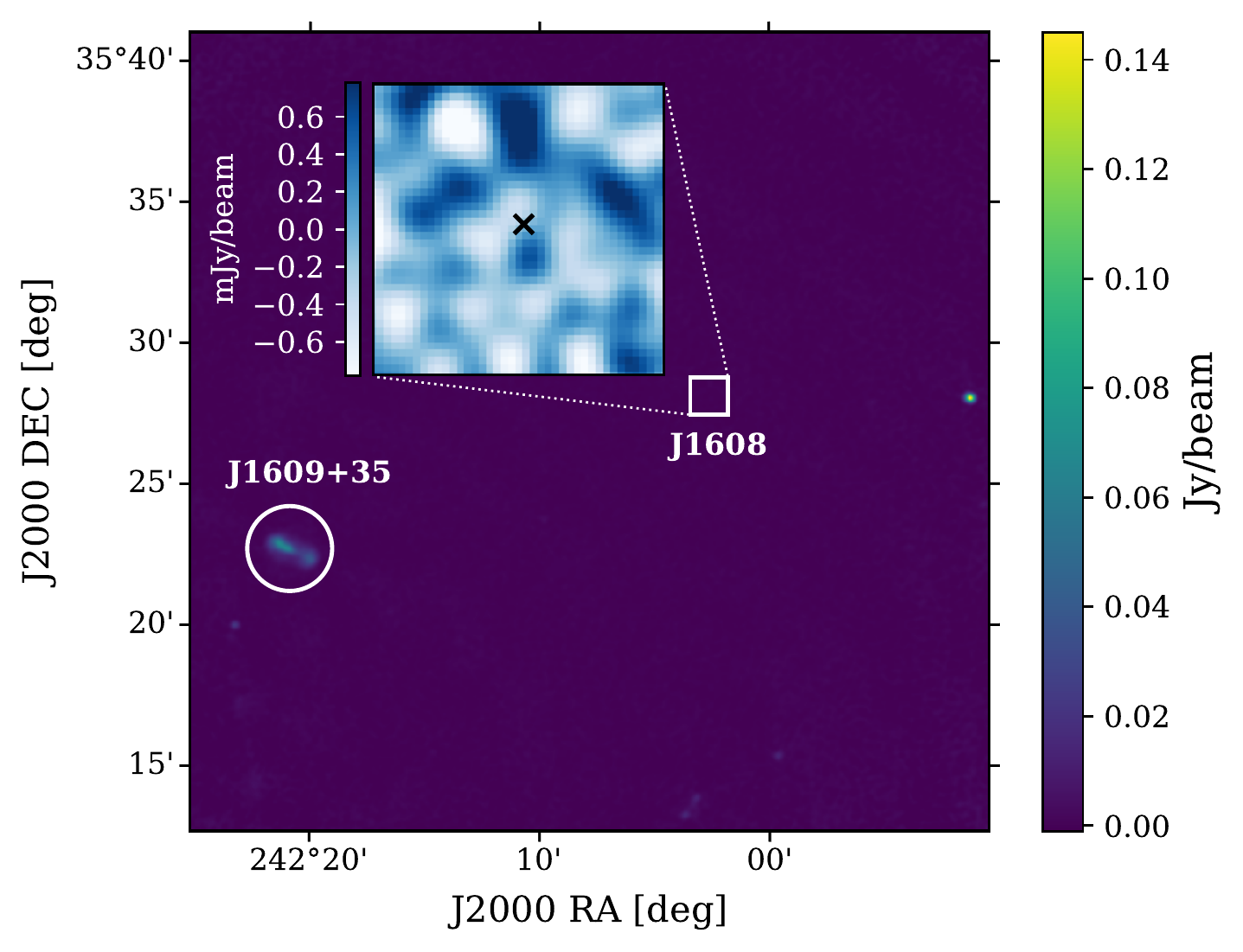}
    \caption{ VLA L-band 20 km/s channel map of the targeted region, centered on J1608's rest frame 21cm frequency at $\nu = 1.3754$GHz.  A white circle and white box indicate positions for J1609+35 and J1608+35. The inset figure shows $\pm3\sigma$ noise fluctuations near J1608's optical coordinates, marked as a black $x$. All other sources in the image are detected in continuum only and show no spectral features. Our non-detection sets an upper limit on J1608's HI mass of $\log M_{HI}/M_\odot<8.4$.}
    \label{stamp}
\end{figure}
\begin{figure*}[tb]
    \centering
    \includegraphics[width=\linewidth]{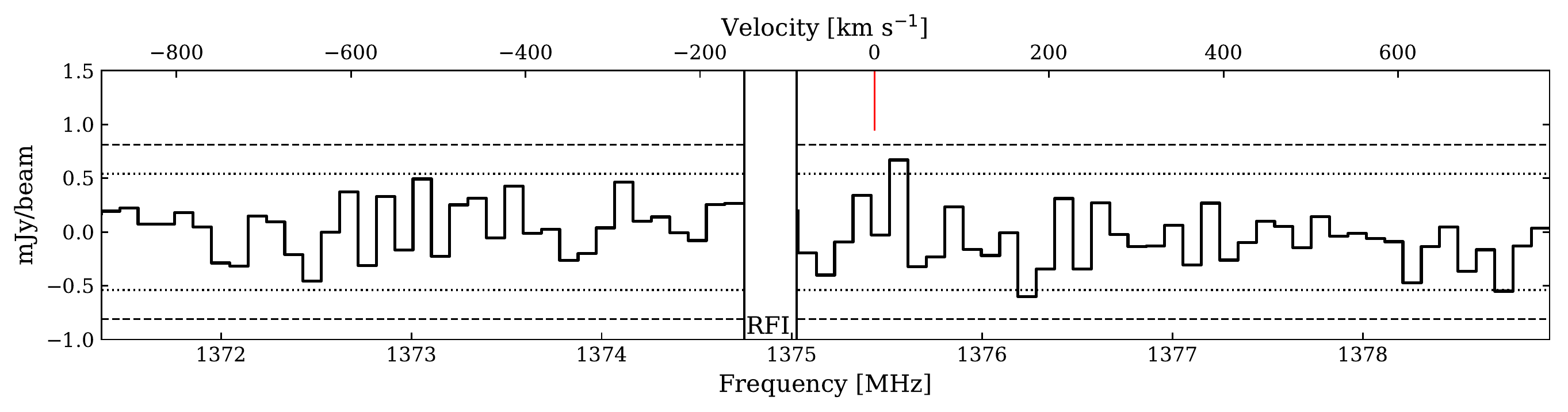}
    \caption{J1608's continuum-subtracted L-Band spectrum taken with the VLA, binned to a resolution of 20 km/s. The top axis shows relative velocities in km/s from J1608's expected restframe 21cm emission for a redshift of $z=0.0327$. Systemic velocity is noted with a vertical red line. Vertical black lines mark an RFI-obscured spectral segment. Horizontal dotted and dashed lines indicate average $\pm2\sigma$ and $\pm3\sigma$ noise uncertainties respectively.}
    \label{21cm}
\end{figure*}
\par In addition to being strong Ly$\alpha$ emitters (LAEs), extreme GPs include some of the only known LyC-leaking star-forming galaxies at low redshift \citep{Izotov2016a,Izotov2016c,Izotov2017,Izotov2018}. As such, GPs are ideal laboratories for studying the escape of ionizing radiation from extreme environments. The high SFR surface densities and emission line strengths seen in GPs indicate large populations of hot, young stars that produce copious ionizing flux (e.g. \citealt{JaskotOey2013,Schaerer2016, Verhamme2017}). At the same time, GPs may have HI column densities $>10^{19}$ cm$^{-2}$, sufficient to absorb LyC photons and scatter Ly$\alpha$ (e.g. \citealt{Gazagnes2018,Chisholm2018}). If feedback clears large cavities of ionized gas, Ly$\alpha$ and LyC photons could escape directly through these channels (\citealt{ClarkeOey2002,Heckman2011}). Thus, the escape properties of GPs may depend on a combination of gas geometry, kinematics and column density. Multi-wavelength studies are required to assess neutral gas signatures such as total covering fraction, outflow velocities and total HI mass. 

Few LCEs have HI observations. Studies of 21-cm emission have been conducted for two confirmed local, LyC-leaking galaxies, Haro 11 and Tololo 1247-232 (hereafter Tol 1247), for which individual star-forming knots were resolved \citep{Pardy2016,Puschnig2017}. Haro 11 has a low HI gas mass of \MHI$/M_*=5.1\times10^8$ \citep{Pardy2016}, and 21cm limits constrain \MHI$/M_*<10^9$ in Tol 1247 \citep{Puschnig2017}. Both have low gas fractions ($M_{HI}/M_*\lesssim0.2$), suggesting that LyC and Ly$\alpha$ escape are related to neutral gas deficiency. However, Tol 1247 and Haro 11 are more massive and leak fewer LyC photons compared to many LCE GPs. 

In this paper we study the relationship between neutral gas and Ly$\alpha$ escape in GPs. We use Very Large Array (VLA) 21cm imaging of the most highly ionized GP J160810+352809 (hereafter J1608), and \textit{Hubble Space Telescope} Cosmic Origins Spectrograph (COS) observations of 17 GPs. These GPs have redshifts $0.02 < z < 0.2$, some of the highest $O_{32}$ ratios ( $\sim 7 - 35$) of SFGs in SDSS and show a variety of Ly$\alpha$ profiles ranging from deep absorption to strong, narrow, double-peaked emission. Ly$\alpha$ escape fractions, defined as the ratio of observed to intrinsic Ly$\alpha$ flux, are as high as 58\%, and Ly$\alpha$ peak separations are as low as $200$ km/s \citep{Henry2015,Jaskot2017}, both potential indicators of low optical depth and LyC escape (e.g. \citealt{Verhamme2017}). With our high-resolution UV spectra, we derive gas column densities and covering fractions and explore their relationship with Ly$\alpha$ emission.

This paper is organized as follows: In section \ref{radio} we present our VLA observations of J1608 and discuss their implications. Section \ref{analysis} describes our UV sample selection and \textit{HST} COS measurements and explores our adopted model. Our UV results are summarized in section \ref{results}. In Section \ref{discussion} we discuss Ly$\alpha$ escape from GPs in the context of high HI column densities and low covering fractions. Section \ref{conclusion} summarizes our conclusions. Throughout this work we adopt a $\Lambda$CDM cosmology with $\Omega_M=0.3$, $\Omega_\Lambda=0.7$ and $H_0=70$ km s$^{-1}$ Mpc$^{-1}$.

\section{21cm VLA Observations of J1608+35}\label{radio}

An enormous $O_{32}=34.9$ and a low Ly$\alpha$ peak separation of 214 km s$^{-1}$ make J1608 a good candidate for escaping LyC radiation (e.g. \citealt{Jaskot2017,Verhamme2017}). J1608 has a low stellar mass $\log M_*/M_\odot=7.04$ (\citealt{Izotov2017}), is roughly $\sim 3'' \times 3''$ in optical SDSS images and has a Petrosian radius ($R_p$), the radius at which the ratio of local surface brightness in an annulus to the mean surface brightness is equal to 0.2, of $R_p=0.3''$ from our \textit{HST} NUV acquisition images. J1608 is extremely ionized and compact, suggesting low neutral gas densities if its HII regions are density-bounded. As mentioned above, local LCEs Tololo 1247-232 and Haro 11 have notably small HI gas fractions ($f_{HI}\leq0.2$) determined from VLA 21cm imaging \citep{Pardy2016,Puschnig2017}, and have likely processed much of their neutral gas reservoir. 

\begin{figure*}
    \centering
    \includegraphics[width=\linewidth]{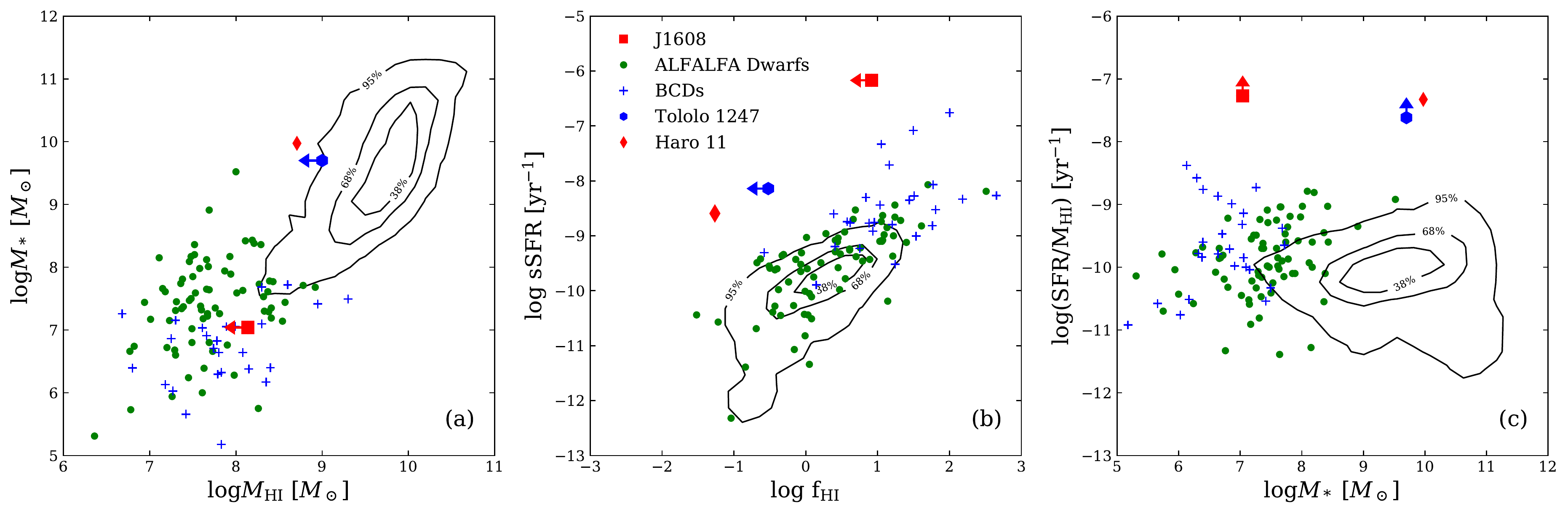}
    \caption{\textit{(a)} Stellar mass vs. HI mass. ALFALFA dwarfs are plotted with a green circle. Blue $+$ markers denote BCDs from \cite{Thuan2016}. J1608 is shown as a red square with an arrow to denote the 3$\sigma$ upper limit. Contours correspond to $38\%,68\%$ and $98\%$ distributions of galaxies in SDSS DR8 \citep{SDSSDR8} with HI masses from the ALFALFA survey \citep{Huang12Scaling}. We also show Tol 1247 and Haro 11 as a red diamond and blue hexagon respectively. \textit{(b)} Specific star-formation rate vs. HI gas fraction ($M_{HI}/M_*$). \textit{(c)} $M_*$ vs. SFR/$M_{HI}$. All SFRs are derived from $H\alpha$, except for ALFALFA dwarfs where we adopt the SFRs presented in \cite{Huang2012}, which may fall below H$\alpha$ estimate by at most $\sim 1$ dex for low SFR. However, this predominantly affects dwarf galaxies at low $f_{HI}$ and does not change the fact that J1608 has an unusually high SFR/$M_{HI}$ ratio. We have scaled the 2$\sigma$ $M_{HI}$ limit for Tololo 1247 from \cite{Puschnig2017} to $3\sigma$ for consistency with J1608. }    
    \label{dwarfs}
\end{figure*}
\subsection{Analysis}

\begin{deluxetable}{llr}[tb]
    \caption{J1608 Predicted HI Mass Fractions\tablenotemark{a}}
    \label{fHI_tab}
    \tablehead{ Correlation  & $\log f_{\mathrm{HI},pre}$ & Reference\tablenotemark{b}}
    \startdata 
    $M_*$  \\
                    & 1.20         & 1 \\[.2ex]
    \hline
    sSFR \\ 
    $\qquad\qquad\mu_*\qquad\qquad$     & 1.56         & 2 \\[.2ex]
    $\qquad\qquad\mu_*$     & 1.65         &1 \\[.2ex]
    $\qquad\qquad\mu_{r90}$ & 1.65         & 1 \\[.2ex]
    \hline
    $NUV-r$ \\ 
    $\qquad\qquad\mu_*$           & 2.62 (1.13)  & 2 \\[.2ex]
    $\qquad\qquad\mu_*$           & 1.39 (1.55)  & 1 \\[.2ex]
    $\qquad\qquad\mu_{r90}$       & 1.36 (1.48)  & 1 \\[.2ex]
    \hline
    $g-r$ \\ 
    $\qquad\qquad\mu_*$             & 2.64 (1.39)  & 1 \\[.2ex]
    $\qquad\qquad\mu_{r90}$         & 2.21 (1.38)  & 1 \\[.2ex]
\enddata

\tablenotetext{a}{Column 1 shows the quantities correlated against each quantity in the subheader whose relation yields the prediction for $\log f_{\mathrm{HI},pre}$ in column 2, using the scaling relations from reference in column 3. Values given in parentheses are estimated using continuum fluxes which exclude line emission. We correct for Milky Way and/or internal extinction as necessary for consistency with each scaling relation.}
\tablenotetext{b}{(1) \cite{Huang12Scaling}, (2) \cite{Zhang2009}}

\label{fHI_table}
\end{deluxetable}
We observed J1608 at 1-2 GHz with the Karl G. Jansky Very Large Array (VLA) under program VLA/16A-176 (PI: Jaskot), targeting rest frame 21cm neutral hydrogen emission. The VLA was in C-configuration (3.5 km maximum baseline) and imaged J1608 for a total of $\sim7.2$ hours on target in three observing sessions in 2016 March and April. For the purposes of self-calibration, we deliberately offset the pointing center by 5 arcmin towards J160923+352242 (hereafter referred to as J1609+35), a 593 mJy continuum source in the NRAO VLA Sky Survey (NVSS; \citealt{NVSS}). All 27 antennas were functional, and we used five spectral windows with a phase center of 1.3754 GHz and individual bandwidths of 2000 kHz. The flux density scale was set by observing the standard flux calibrator J1331+3030 (3C286) at the start of each night. Phase calibrations were determined by observing J1613+3412 at regular intervals. 

We manually flagged for radio frequency interference (RFI) and faulty baselines using the Common Astronomy Software Applications package (CASA; \citealt{CASA}) version 5.0.0. A known RFI signature $\sim150$ km/s from the target frequency appeared in each observing epoch and was present in visibility data for the science goal and each calibrator. The corresponding channels were removed entirely to minimize noise near expected 21cm emission. Prior to calibration, we masked visibilities at potential spectral line frequencies from down-weighting. We then calibrated each night with the CASAv4.5.0 automated reduction pipeline, turning off Hanning smoothing to preserve the spectral resolution. Separate observations were re-gridded to the same velocity axis and combined prior to continuum subtraction to enhance emission from J1608 if present. However, no continuum emission was detected in individual nights or the combined dataset. Continuum emission from other sources in the field were linearly fit in the $uv-$plane and subtracted from the visibility data. To image and deconvolve, we interactively cleaned using \texttt{tclean} with a default gain setting of 0.1. We used Briggs weighting and set the robust parameter $R=0.5$ to balance between resolution and sensitivity. The restoring beam size is $\sim 16'' \times 14''$ with a position angle of -73.8 degrees. Figure \ref{stamp} shows a slice of the VLA spectral cube at J1608's redshifted 21cm frequency of $1.3754$GHz after binning the data to a resolution of 20 km/s. We measure a continuum flux density for J1609+35 of $607.9\pm0.22$ mJy. 

The 21cm hydrogen hyperfine structure line was not detected in J1608's VLA spectrum, as shown in Figure \ref{21cm}. We place an upper limit on the total HI gas mass using an RMS of 0.261 mJy/beam estimated from the VLA map binned to 20 km/s. J1608 is unresolved in the VLA beam, corresponding to a flux density limit of 0.261 mJy. We set an HI mass upper limit using the relation of \cite{Roberts1994}
\begin{equation}
    \frac{M_{HI}}{M_\odot} = 2.36\times10^5 \Big(\frac{D}{\mbox{Mpc}}\Big)^2
    \Big(\frac{S_{21cm}}{\mbox{Jy}}\Big)
    \Big(\frac{W_{21cm}}{\mbox{km/s}}\Big) 
\end{equation}
where $D$ is the luminosity distance, $S_{21cm}$ is the 21cm line flux, and $W_{21cm}$ is the 21cm line width. We assume a 3$\sigma$ detection threshold of $S_{21cm}\approx0.78$ mJy and $W_{21cm}=36.1$ km/s, characteristic of HI dwarf galaxies in the The Arecibo Legacy Fast ALFA (ALFALFA) survey with stellar masses between $10^6 - 10^8$ $M_\odot$ \citep{Huang2012}. We place a $3\sigma$ upper limit on J1608's HI mass at $\log M_{HI}/M_\odot \leq 8.14$.

\subsection{Comparison with other Galaxies}

J1608's neutral gas mass is either typical or below average for galaxies of comparable stellar mass. Figure \ref{dwarfs}a compares J1608's stellar and HI masses with those of blue compact dwarfs (BCDs) \citep{Thuan2016}, and HI-selected ALFALFA dwarfs \citep{Huang2012}. Our upper limit on J1608's HI mass is near the average of both comparison dwarf populations. 

To further investigate the HI content of J1608 we calculate a $3\sigma$ upper limit on its HI mass fraction ($f_{HI}\equiv M_{HI}/M_*$) of $\log f_{HI}\leq1.1$ assuming $\log M_*/M_\odot =7.04$ \citep{Izotov2017}. We then compare $f_{HI}$ with predicted HI mass fractions ($f_{HI,pre}$) from scaling relations using combinations of optical and UV color, $M_*$, stellar mass surface density $\mu$, and sSFR. To estimate the stellar mass surface density, we adopt J1608's stellar mass reported in \cite{Izotov2017}. We then calculate $\mu_*$ and $\mu_{r90}$ consistently with the published scaling relations as $\mu_*=0.5M_*/(\pi r_{50z}^2)$ and $\mu_{r90}=0.5M_*/(\pi r_{90r}^2)$ where $r_{50z}$ is the radius containing 50\% of Petrosian flux in the SDSS \textit{z} band, and $r_{90r}$ the radius containing 90\% of Petrosian flux in the SDSS \textit{r} band.

We use scaling relations calibrated on optically-selected SDSS galaxies \citep{Zhang2009} and the HI-selected ALFALFA sample \citep{Huang12Scaling}. In general, these scaling relations are linear over four orders of magnitude in $f_{HI}$ with typical RMS scatter of $\sigma\sim0.3$ dex. We correct J1608's observed magnitudes for Milky Way and/or internal extinction as necessary for consistency with each scaling relation. To correct for Milky Way extinction, we use the \cite{fitzpatrick99} extinction law and \cite{schlafly11} dust map. We correct J1608's $NUV-r$ and $g-r$ colors for internal extinction when necessary using techniques discussed in Section \ref{analysis}. In addition to using observed magnitudes, we also estimate $g,r$ and $NUV$ from optical continuum fluxes as colors could be skewed by J1608's strong metal line emission. We estimate line-free magnitudes by convolving continuum fits through the SDSS filters. For consistency with comparison samples, we use J1608's dust-corrected H$\alpha$ luminosity of 9.473$\times10^{41}$ erg s$^{-1}$ and the \cite{Kennicutt1998} SFR relation to estimate $\log$ SFR$_{H\alpha}/(M_\odot\mbox{ yr}^{-1})=0.876$. The SFR derived from SED modeling is 0.6 $M_\odot$/yr \citep{Izotov2017}, close to our H$\alpha$-derived value.

Table \ref{fHI_tab} lists $f_{\mathrm{HI},pre}$ calculations for multiple scaling relations. Our VLA-derived \FHI\ upper limit is below all predictions by a factor of $\sim0.5$ dex on average. Adopting an HI line-width of $W_{21cm}\sim60$ km/s boosts J1608's HI limits to $\log M_{HI}/M_\odot \leq 8.36$ and $\log f_{HI}\leq1.3$, which remains below the majority of predicted HI mass fractions. The scaling relations used in estimating $f_{\mathrm{HI},pre}$ are independent of HI assumptions. Moreover, $\sim15\%$ of dwarf galaxies in the ALFALFA sample with stellar masses between $\log M_*/M_\odot=6.5-7.5$ derived from SED fitting have HI line widths $>60$ km/s \citep{Huang2012}. Thus, an HI line width $>60$ km/s is unlikely, and J1608 has less neutral gas than would otherwise be expected for its bright UV/optical photometry.

While J1608 may be typical in terms of its baryonic mass alone, the SF properties of this GP are highly unusual. As is the case for Tololo-1247 and Haro 11, J1608's H$\alpha$-derived sSFR is nearly an order of magnitude greater than galaxies of comparable gas fraction (Fig. \ref{dwarfs}b). However, star-formation in dwarf galaxies is not constant (e.g. \citealt{McQuinn2010b,Weisz2011,Hopkins2014}); J1608's sSFR estimate may be enhanced by a recent burst. A young starburst would boost J1608's H$\alpha$ emission and therefore its H$\alpha$-derived SFR. A recent burst could also explain J1608's unusual SFR/$M_{HI}$ for its $M_*$ (Fig. \ref{dwarfs}c). J1608, Tololo-1247 and Haro 11 all likely have unusually large quantities of massive stars relative to their HI masses. Stellar feedback, either radiative or mechanical, may overpower the neutral gas content in these galaxies more easily, clearing optically thin channels by which LyC and Ly$\alpha$ photons can escape.

\section{UV Spectra of extreme green peas}\label{analysis}
\begin{deluxetable*}{lcccccccc}
    \caption{Green Pea sample properties. }
    \label{props}
    \tablehead{Galaxy & $z$ & 12+$\log$(O/H) & $O_{32}$ & $f_{esc}^{Ly\alpha}$ & EW(Ly$\alpha$)\tablenotemark{a} & $\Delta v_{\mathrm{Ly}\alpha}$\tablenotemark{b}  & EW(H$\alpha$) & $A_V$ \\ & & & & & (\AA) & (km s$^{-1}$)& (\AA) &  }
    \startdata 
    J144805-011058 & 0.0274 & 8.11$^{+0.04}_{-0.05}$ & 7.8$\pm0.3$  & 0.0             & $-18\pm0.1$ & \---  & $ 805 \pm 6       $ & $0.36\pm0.03$\\[.2ex]
    J160810+352809 & 0.0327 & 7.83$^{+0.13}_{-0.19}$ & 34.9$\pm3.5$ & $0.18 \pm 0.04$ & $163\pm12$ & $214\pm30$ & $ 1472 \pm 23 $ & $0.39\pm0.05$\\[.2ex]
    J133538+080149 & 0.1235 & 8.10$^{+0.22}_{-0.45}$ & 7.3$\pm$0.4  & 0.0             & $-14\pm0.5$ & \---  & $ 827 \pm 9       $ & $0.14\pm0.04$\\[.2ex]
    J145735+223202 & 0.1487 & 8.05$^{+0.09}_{-0.12}$ & 7.2$\pm$0.5  & $0.01 \pm 0.01$ & $-4\pm1$  & $749\pm56$ & \---             & $0.24\pm0.03$\\[.2ex]
    J150934+373146 & 0.0326 & 7.88$^{+0.08}_{-0.10}$ & 15.1$\pm0.9$ & $0.05 \pm 0.02$ & $12\pm1$ & $400\pm27$ & $ 1411 \pm 14   $ & $0.24\pm0.04$\\[.2ex]
    J085116+584055 & 0.0919 & 7.87$^{+0.10}_{-0.14}$ & 9.4$\pm$0.5  & $0.04 \pm 0.01$ & $26\pm2$ & $361\pm25$ & $ 1595 \pm 23   $ & $0.36\pm0.04$\\[.2ex]
    J021307+005612 & 0.0399 & 8.03$^{+0.08}_{-0.10}$ & 7.2$\pm0.4$  & 0.12 $\pm 0.02$ & $42\pm4$ & $397\pm47$ & $ 1016 \pm 10   $ & $0.42\pm0.04$\\[.2ex]
    J122612+041536 & 0.0942 & 7.99$^{+0.10}_{-0.12}$ & 8.3$\pm$0.5  & $0.13 \pm 0.02$ & $64\pm3$ & $360\pm40$ & $ 1060 \pm 14   $ & $0.28\pm0.04$\\[.2ex]
    J024052-082827 & 0.0822 & 7.91$^{+0.09}_{-0.12}$ & 13.7$\pm$0.6 & $0.19 \pm 0.04$ & $154\pm8 $ & $266\pm29$ & $ 1752 \pm 17 $ & $0.33\pm0.03$\\[.2ex]
    J173501+570309 & 0.0472 & 8.11$^{+0.07}_{-0.08}$ & 6.8$\pm0.3$  & $0.09 \pm 0.04$ & $64\pm4$& $460\pm47$ & $ 1442 \pm 10    $ & $0.41\pm0.02$\\[.2ex]
    J230210+004939 & 0.0331 & 7.72$^{+0.07}_{-0.08}$ & 8.6$\pm0.6$  & $0.28 \pm 0.06$ & $64\pm3$& $279\pm48$ & $ 897 \pm 11     $ & $0.16\pm0.04$\\[.2ex]
    J131131-003844 & 0.0811 & 7.98$^{+0.12}_{-0.17}$ & 6.6$\pm$0.2  & $0.23 \pm 0.05$ & $71\pm4$ & $273\pm26$ & $ 1106 \pm 10   $ & $0.36\pm0.02$\\[.2ex]
    J120016+271959 & 0.0819 & 8.05$^{+0.06}_{-0.07}$ & 8.9$\pm$0.5  & $0.39 \pm 0.08$ & $114\pm7$ & $327\pm65$ & $ 1057 \pm 10  $ & $0.19\pm0.03$\\[.2ex]
    J080841+172856 & 0.0442 & 7.61$^{+0.13}_{-0.18}$ & 10.3$\pm$0.9 & $0.36 \pm 0.07$ & $31\pm2$ & $146\pm37$; $441\pm58$\tablenotemark{c} & $ 424 \pm 5$ & $0.35\pm0.03$\\[.2ex]
    J081552+215624 & 0.1410 & 8.02$^{+0.12}_{-0.03}$ & 10.1$\pm$0.7 & $0.28\pm0.06$   & $68\pm4$ & $296\pm52$ & \---              & $0.13\pm0.04$\\[.2ex]
    J030321-075923 & 0.1649 & 7.91$^{+0.12}_{-0.12}$ & 7.1$\pm$0.5  & $0.05\pm0.01$   & $6\pm1$ & $443\pm156$& $697\pm12$         & $0.03\pm0.04$\\[.2ex]
    J121904+152609 & 0.1956 & 7.88$^{+0.21}_{-0.03}$ & 10.5$\pm$0.7 & $0.58\pm0.08$   & $17\pm9$ & $242\pm43$ & $1266\pm21$       & $0.09\pm0.04$\\[.2ex]
 \enddata
 \tablenotetext{a}{Measurement includes both absorption and emission. Negative values indicate net absorption. See \cite{Jaskot18} for details. }
 \tablenotetext{b}{Ly$\alpha$ peak separation from \cite{Jaskot2017}.}
 \tablenotetext{c}{Triple-peaked system with two blue Ly$\alpha$ peaks. }
\end{deluxetable*}

\begin{figure*}[tb]
    \centering
    \includegraphics[width=\textwidth]{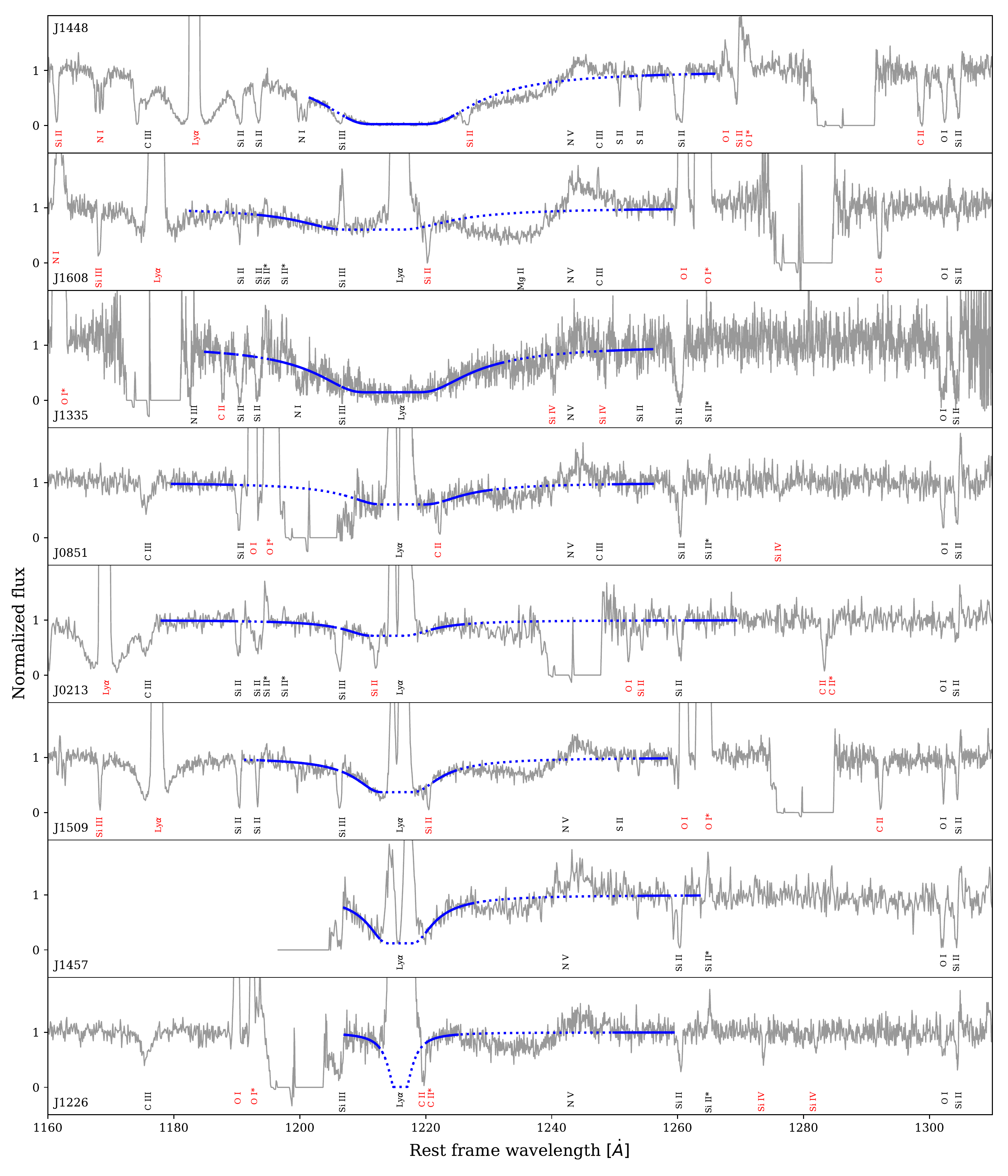}
    \caption{COS spectra of GPs showing Ly$\alpha$ absorption. Targets are sorted in order of decreasing HI column density from top to bottom. Best-fit models are shown in solid blue. Dotted segments are removed from the fitting routine due to the presence of Milky Way, stellar or nebular features. Milky Way absorption lines are identified in red and GP lines in black. }
    \label{allspecs}
\end{figure*}

\begin{figure*}[tb]
    \centering
    \includegraphics[width=\textwidth]{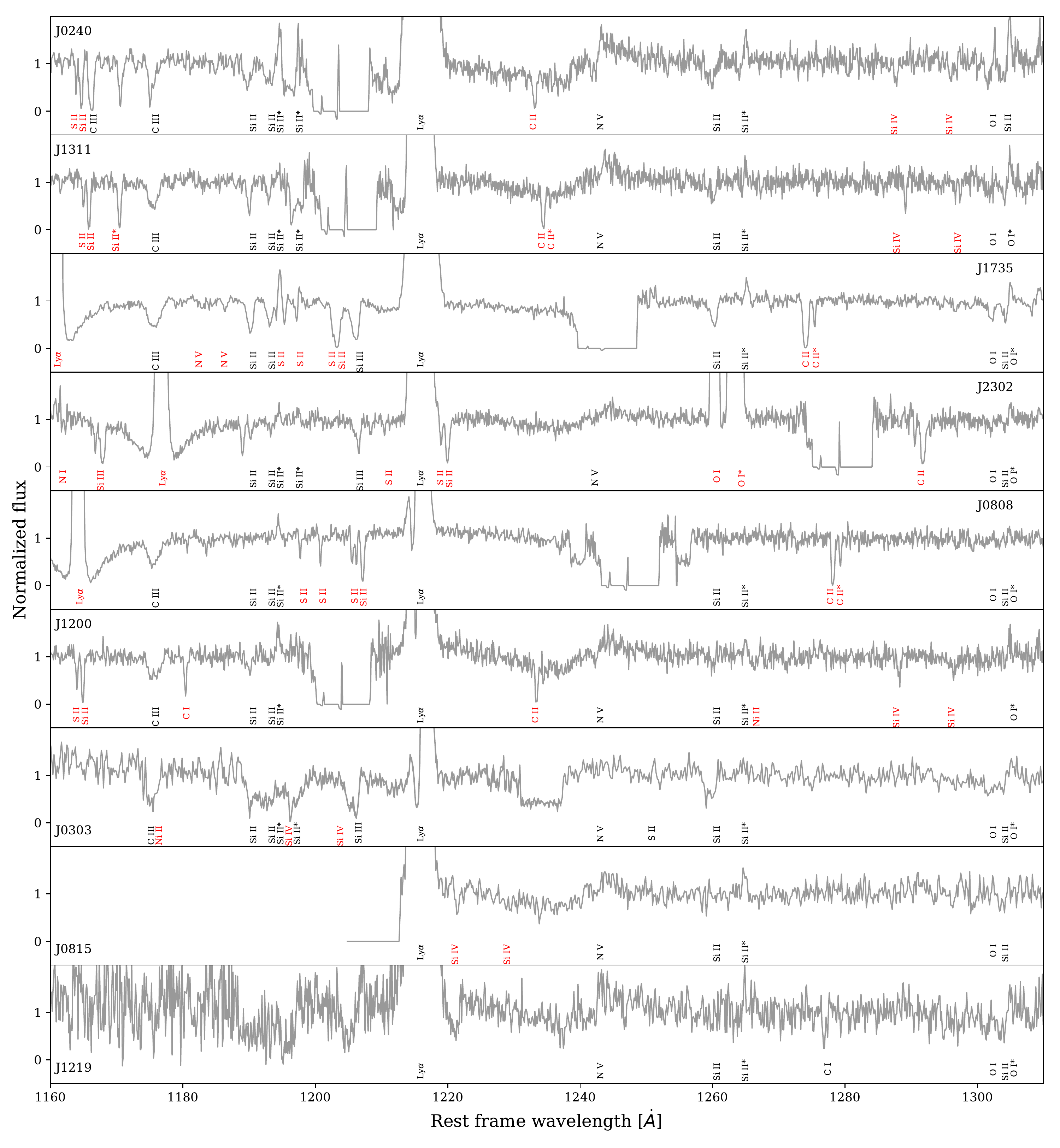}
    \caption{Same as figure \ref{allspecs} but showing GPs for which models could not be implemented due to weak or absent Ly$\alpha$ absorption.}
    \label{nomods}
\end{figure*}

We analyze \textit{HST} COS spectra of 17 GPs, of which 13 are from Program GO-14080 (PI Jaskot) and were observed with the G130M grating in Cycle 23 at lifetime Position 3. \cite{Jaskot18} describes the \textit{HST} program in more detail, and also includes SDSS and COS NUV acquisition images of the sample. J145735+223202 (hereafter J1457) and J081552+215624 (hereafter J0815) were observed during Program GO-13293 (PI Jaskot) with the G160M grating. J121904+152609 (hereafter J1219) and J030321-075923 (hereafter J0303) are from GO-12928 (PI Henry) and were observed with the G130M and G160M grating. Our cumulative sample spans a redshift range of $z=0.0274-0.1956$ and was selected for large \Orat ratios and high S/N from the SDSS Data Release 10 \citep{SDSSDR10}. The wavelength coverage of each spectrum is $\Delta\lambda \approx 300$ \AA\ with a range common to most of $\lambda_{rest} = 1120-1340$ \AA, always including redshifted Ly$\alpha$. Data reduction techniques are described in (\citealt{Jaskot18}) and are summarized here. 

Each COS observation was binned to its own spectral resolution depending on the spatial extent of the target. Resolutions for sources observed in GO-14080 are between $12-34$ km/s, marginally wider than point source profiles and are derived from the FWHM of the cross-dispersion profiles. GPs from GO-13293 and GO-12928 are binned to resolutions from $30-41$ km/s and $28-45$ km/s respectively to increase S/N. Separate resolutions were calculated for the A and B spectral segments. Linear continuum fits were estimated in the $1140-1290$ \AA\ rest-frame region after masking out absorption and emission features. Milky Way extinction was accounted for using the \cite{fitzpatrick99} law and \cite{schlafly11} extinction maps. As discussed in \cite{Jaskot2017} and \cite{Jaskot18}, the UV continuum and Ly$\alpha$ emission trace comparably compact regions, and Ly$\alpha$ emission is not significantly more extended than the UV continuum in the GPs' COS 2D spectra. Thus, the GPs' Ly$\alpha$ emission likely originates from the starburst region itself.

We used optical SDSS emission lines to determine the GPs' internal reddening, oxygen abundances, and \Orat ratios. To correct for internal extinction, we follow \cite{Izotov2017} and use the H$\alpha$/H$\beta$ ratios and the \cite{Cardelli1989} extinction law, which was found to provide better fits to UV/IR fluxes and emission lines in extreme emission line galaxies (see also \citealt{Jaskot18}). We adopt $R_V=2.7$ in GPs with H$\beta$ EWs $>150$\AA\ and $R_V=3.1$ otherwise.

We calculate Ly$\alpha$ escape fractions as in \cite{Jaskot2017}, using the dust-corrected H$\alpha$ fluxes and the Ly$\alpha$/H$\alpha$ ratios appropriate for each GP's temperature. Electron temperatures were calculated using \texttt{PyNeb} \citep{pyneb} with the dust-corrected $\lambda\lambda5007,4959$ to $\lambda4363$ flux ratios. Oxygen abundances were also calculated in \texttt{PyNeb} via the direct method, using the derived electron temperatures, [O III]$\lambda5007,4959$ fluxes and [O II] $\lambda3272$ fluxes. We adopt an ionization correction factor for non-detected O ionization states from \cite{PerezMontero2017}, constrained by the GPs' HeI and HeII emission. Ly$\alpha$ emission properties, metallicities, redshifts and \Orat values are reported in Table \ref{props}. 

\begin{figure}[tb]
    \centering
    \includegraphics[width=\linewidth]{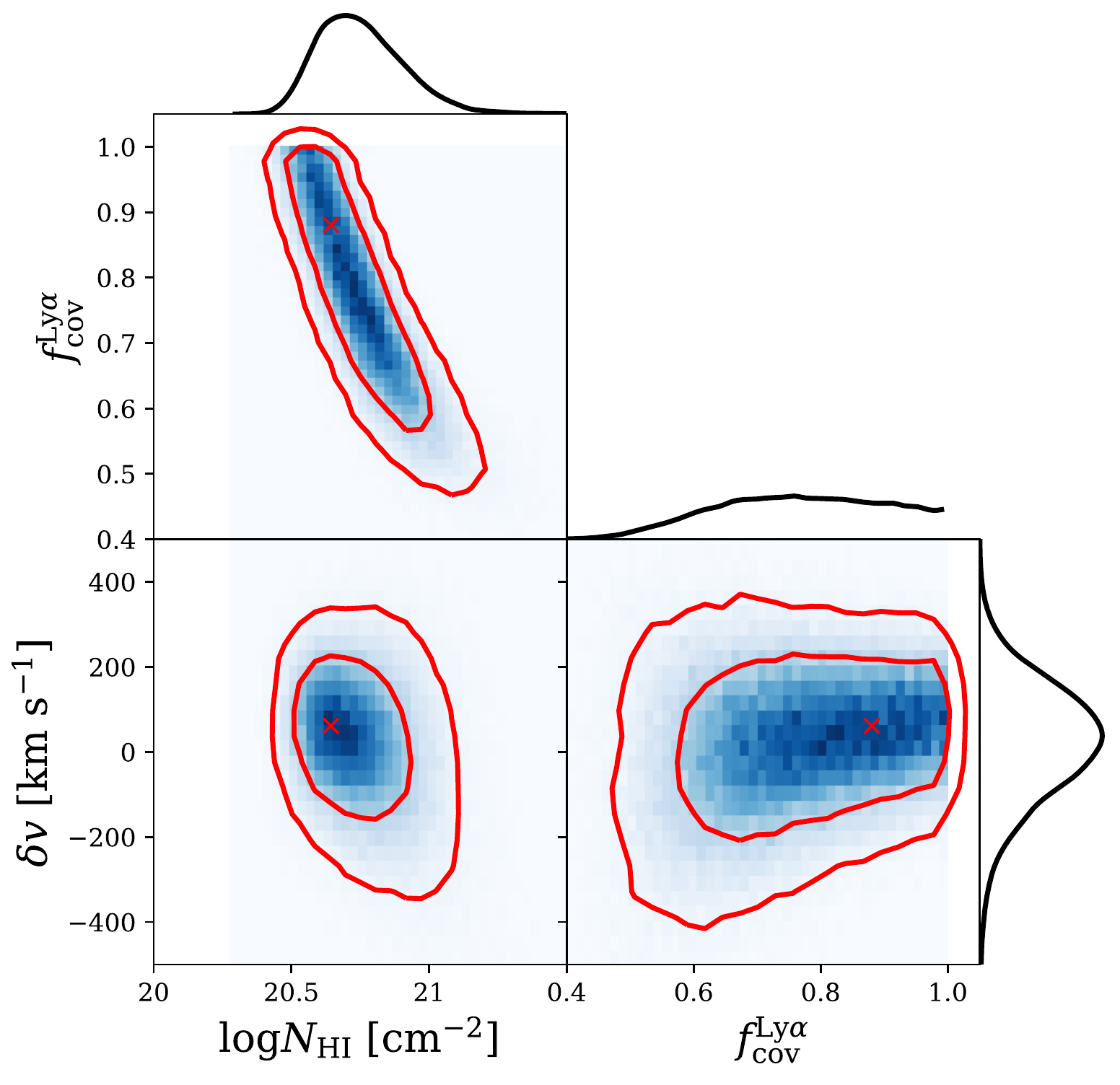}
    \caption{Example joint posterior distributions for \NHI, $\delta v$ and \fcovLyA taken from the fit to J1457. The Doppler parameter $b$ has a flat posterior in all fits and is not shown. Red contours represent 68\% and 99\% confidence limits. The red cross indicates best-fit parameter values where the likelihood function is maximized. Posteriors are drawn in black on their respective axes. }
    \label{posts}
\end{figure}

\subsection{\NHI\ Derived from Ly$\alpha$}\label{LyA_N}

Eight out of 17 GPs show significant Ly$\alpha$ absorption wings, an unusual statistic compared to previous studies of GPs with lower [O III]/[O II]. For instance \cite{Yang2017} found non-negligible Ly$\alpha$ absorption in only $1/3$ of 48 GPs. Furthermore \cite{Henry2015} and \cite{Verhamme2017} did not detect any Ly$\alpha$ absorption in a sample of 10 GPs and 5 GP LCEs respectively. We note that different integration times between samples and low S/N may affect the detection of Ly$\alpha$ absorption. 

Six of the GPs in our sample that show Ly$\alpha$ absorption also show strong, double-peaked Ly$\alpha$ emission. High Ly$\alpha$ escape fractions and low Ly$\alpha$ peak separation make some of these GPs good LCE candidates \citep{Verhamme2015}, and the presence of both Ly$\alpha$ absorption and emission in the same galaxy suggests both optically thick and optically thin regions along the line of sight. This geometry may be consistent with a scenario in which Ly$\alpha$ and LyC escape through low-column density channels (e.g. \citealt{Heckman2011,RivThorsen2015,Puschnig2017,Gazagnes2018,Chisholm2018}). 

We measure HI column densities for the eight Ly$\alpha$-absorbing GPs in our sample by fitting Voigt profiles to the Ly$\alpha$ absorption. The radiative transfer equation for pure absorption is  
\begin{equation}
F(\lambda|N,b,z) = F_0 (\lambda) e^{-\tau(\lambda|N,b,z)}
\label{flambda}
\end{equation}
where $F$ is the emergent flux, $F_0$ is the continuum, $N$ is the absorbing column density, $b$ is the Doppler parameter and $z$ is the target's redshift. Voigt profiles are the convolution of a Gaussian and Lorentzian profile, with optical depth
\begin{equation}
\tau(\lambda|N,b,z) = N \sigma_0 f \Phi(\lambda|b,z)
\label{tau}
\end{equation}
where $\Phi(\lambda)$ is calculated with the real part of the Faddeeva function as implemented in \texttt{SciPy}. The cross section and line oscillator strength are given by $\sigma_0$ and $f$ respectively. We allow $N$ and $b$ to vary in our fits but keep $z$ fixed at its spectroscopically confirmed value. Ly$\alpha$ line centers are allowed to vary between $\pm1000$ km/s to account for bulk gas motion through a velocity shift parameter $\delta v$. \cite{Jaskot2017} reported evidence for low outflow velocities $<300$ km/s in this sample, and in practice the majority of fits never reach such extreme values. 

Observational evidence suggests that absorbing gas may not fully cover ionizing sources in some star-forming galaxies (e.g \citealt{Heckman2011,RivThorsen2015,Gazagnes2018}). We consider this scenario by using a covering fraction parameter \fcovLyA, and adopting the ``picket-fence" model with a uniform dust screen introduced by \cite{Heckman2001}. The intensity at the core of the Ly$\alpha$ line may be non-zero either due to non-uniform covering fraction (e.g. \citealt{Heckman2011}) or due to infilling, where hydrogen gas scatters Ly$\alpha$ photons into the line of sight. The final spectral model is
\begin{multline}
F(\lambda|N,b,\delta v,\mathcovLyA) = F_0(\lambda) \times \\\Big(\mathcovLyA  F_{Voigt}(\lambda|N,b,\delta v) + (1-\mathcovLyA)\Big)
\label{model}
\end{multline}

We stress that our model is intended to only fit damped Ly$\alpha$ absorption wings in low-dust GPs. We exclude all nebular, stellar and Milky Way lines from the fit as well as central Ly$\alpha$ emission if present. We note that the above model is commonly modified to allow for unattenuated stellar emission emerging from optically thin regions (e.g. \citealt{Zackrisson2013,Gazagnes2018}). However, our sample has $A_V=0.03-0.42$ for which the ratio in model-predicted flux with and without dust porosity is on the order of one part in $10^{5}$. Therefore, the assumption of a particular dust geometry does not significantly impact our results. Finally, we checked that stellar Ly$\alpha$ absorption is negligible by comparing against the Binary Population and Spectra Synthesis (BPASS) stellar population models (BPASSv2.1; \citealt{BPASS}). For ages $< 5$ Myr, as expected for the GPs (e.g. \citealt{JaskotOey2013}), stellar Ly$\alpha$ absorption profiles do not extend beyond $\pm10$ km/s of line center whereas the absorption seen in the COS spectra appears beyond $\pm1000$ km/s from line center. 

We fit the Ly$\alpha$ absorption wings using Markov Chain Monte Carlo (MCMC). We use the stretch-move algorithm as implemented in \texttt{emcee}, an open-sourced MCMC routine presented in \cite{emcee}. We use 100 walkers that move as an ensemble, taking 10,000 steps through the 4-dimensional parameter space. We assume uniform priors with consistent limits across all fits. The fit's output is a set of posterior probability distributions for each parameter. 

Our derived HI column densities are listed in Table \ref{all_columns}. We determine best-fit values from the MCMC step that maximizes the likelihood function. In the eight modeled GPs, we measure HI column densities greater than $10^{19}$ cm$^{-2}$, evidence for significant neutral gas reservoirs along the line of sight, even in targets with strong Ly$\alpha$ emission. Figure \ref{allspecs} shows the model fits and illustrates the strong and often double-peaked Ly$\alpha$ emission profiles superimposed on top of deep absorption. In comparison, we show all other GP spectra for which no models could be fit in Figure \ref{nomods}. We note that GPs such as J0240 and J1311 still show Ly$\alpha$ absorption, but the COS chip-gap prevents us from fitting a reliable model. 

Figure \ref{posts} shows an example of the typical free-parameter covariances and behaviors seen in our fits. The Voigt Doppler parameter $b$ has no impact on the goodness-of-fit in every case. We find that \fcovLyA\ posteriors flatten out after an initial peak, are overall poorly constrained but do influence HI column density measurements. Best-fit \NHI\ values increase by $\sim0.3$ dex on average when compared to models with fixed $\mathcovLyA=1$. In either case, \NHI\ posteriors are generally Gaussian in appearance with standard deviations on the order of $\leq 0.1$ dex for a given value of \fcovLyA. 

Column density and velocity offset are typically un-correlated in our sample except for J1608, J0213 and J0851, where Ly$\alpha$ absorption is obscured by either the COS chip-gap or broad N V P-Cygni features seen in young, massive stars over the $\sim1220\mathrm{\AA}\--1240\mathrm{\AA}$ range. In such cases the MCMC algorithm finds good fits by either increasing column density or shifting the entire Voigt profile blueward or redward. Small velocity offsets $\lesssim40$ km/s seen in low-ionization metal lines suggest that for these objects a higher column density scenario is more likely (see \citealt{Jaskot2017}). We re-fit J1608, J0213 and J0851 with fixed $\delta v$ and find that best-fit column densities change by only 0.1\%. HI column densities and errors presented for J0213 and J0851 in Table \ref{all_columns} correspond to models allowing $\delta v$ to vary. We fix $\delta v=0$ when fitting J1608, a reasonable assumption as most of J1608's metal lines show nearly zero velocity (e.g. \citealt{Jaskot2017}). 

\begin{deluxetable*}{lcllllll}
    \tabletypesize{\small}
    \caption{Derived Column Densities and Covering Fractions.}
    \label{all_columns}
    \tablehead{GP & $\log$\NHI/cm$^{-2}$ & $\log N$/cm$^{-2}$ & $\log N$/cm$^{-2}$ & $\log N$/cm$^{-2}$ & $\log N$/cm$^{-2}$ & $\log N$/cm$^{-2}$ & $\mathcov^{\mbox{\scriptsize SiII}}$ \\
     & Ly$\alpha$ &  Si II $\lambda1190$ & Si II $\lambda1193$ & Si II $\lambda1260$& Si II $\lambda1304$ & O I $\lambda1302$ & } 
    \startdata 
    J1448-01 & $21.49 \pm 0.02$ & $14.34\pm 0.23$   & $14.16 \pm 0.33$  & $14.17\pm0.16$ & $14.32\pm0.12$ & $15.27\pm 0.24$ & $ 0.94 \pm 0.03 $\\[.2ex]
    J1608+35 & $21.39 \pm 0.13$ & $14.03 \pm 0.19$  & $13.64 \pm 0.21$  & -              & $13.95\pm0.30$ & $14.05\pm0.20$& $ 0.42 \pm 0.08 $  \\[.2ex]
    J1335+08 & $21.23 \pm 0.05$ & $15.60\pm 0.17$   & $14.64 \pm 0.28 $ & $14.79\pm0.17$ & $15.48\pm0.42$ & $15.63\pm0.31$& $ 0.97 \pm 0.04 $  \\[.2ex]
    J1457+22 & $20.55 \pm 0.11$ & -                 & -                 & $14.07\pm0.42$ & $14.95\pm0.47$ & $14.88\pm0.56$& $ 0.99 \pm 0.12 $  \\[.2ex]
    J1509+37 & $20.49 \pm 0.03$ & $14.49 \pm 0.21$  & $13.99 \pm 0.18$ & -              & $14.55\pm0.17$ & $14.79\pm0.07$& $ 0.82 \pm 0.03 $  \\[.2ex]
    J0851+21 & $20.35 \pm 0.07$ & $14.49 \pm 0.31$ & -                 & $14.33\pm0.49$ & $14.65\pm0.41$ & $14.96\pm0.38$ & $ 0.86 \pm 0.04 $ \\[.2ex]
    J0213+00 & $20.00 \pm 0.21$ & $14.27 \pm 0.08$ & $14.12 \pm 0.12 $ & $13.61\pm0.27$ & $14.27\pm0.17$ & - & $0.54 \pm 0.03 $             \\[.2ex]
    J1226+04 & $19.47 \pm 0.64$ &   -               & -                 & $13.59\pm0.31$ & $14.38\pm0.40$ & $14.64\pm0.31$ & $ 0.60 \pm 0.03 $ \\[.2ex]
    J0240-08 & -                & $14.29 \pm 0.19$  & $13.96 \pm 0.29$  & $13.55\pm0.23$ & -              & -          & $ 0.38 \pm 0.08 $     \\[.2ex]
    J1735+57 & -                & $14.45 \pm 0.40$  & $14.01 \pm 0.10$  & $13.64\pm0.20$ & -              & $14.66\pm0.13$ & $ 0.49 \pm 0.04 $ \\[.2ex]
    J2302+00 & -                & $13.91 \pm 0.28$  & -                 & -              & -              & -      & $ 0.28 \pm 0.12 $          \\[.2ex]
    J1311+00 & -                & $14.21 \pm 0.22$  & $13.58\pm  0.16$  & $13.26\pm0.20$ & $14.30\pm0.19$ & -   & $ 0.45 \pm 0.02 $            \\[.2ex]
    J1200+27 & -                & $13.92 \pm 0.21$  & -                 & $12.95\pm0.24$ & -              & -       & $ 0.24 \pm 0.10 $        \\[.2ex]
    J0808+17& -& $<12.04$          & $<11.70$          & $<11.16$       & $<12.33$       & $<12.52$    & $<0.16$    \\[.2ex]
    J0815+58               & -& -                 & -                 & $<11.16$       & $<12.43$       & $<12.58$  &  $<0.45$     \\[.2ex]
    J0303-07                & -& -                 & -                 & $13.78\pm0.49$ & $14.90\pm0.42$ & $15.12\pm0.46$  & $0.40\pm0.07$ \\[.2ex]
    J1219+15            & -& $<12.52$          & $<12.21$          & $<11.60$       & $<12.77$       & $<12.94$   &  $<0.93 $   \\[.2ex]
 \enddata
 \tablecomments{GPs are ordered by decreasing $\log$\NHI. Si II and O I column densities were calculated using the apparent optical depth method for all available absorption lines in each target. Upper limits on column densities and covering fractions were calculated for GPs without detected LIS absorption, and all upper limits are quoted at $3\sigma$ levels.}
\end{deluxetable*}

\begin{deluxetable*}{lccccc}
    \caption{Derived Equivalent Widths.}
    \label{ews}
    \tablehead{GP  & EW(Si II $\lambda1190$) & EW(Si II $\lambda1193$) & EW(Si II $\lambda1260$)& EW(Si II $\lambda1304$) & EW(O I $\lambda1302$) \\ 
    & (\AA) &(\AA) &(\AA) &(\AA) & (\AA)} 
    \startdata 
    J1448-01 & $ 1.15 \pm 0.19 $ & $ 0.84 \pm 0.09 $  & $ 1.39 \pm 0.13 $ & $ 0.7 \pm 0.1 $   & $ 0.94 \pm 0.15 $  \\[.2ex]
    J1608+35 & $ 0.17 \pm 0.09 $ & $ 0.04 \pm 0.04 $  & -                 & $ 0.19 \pm 0.12 $ & $ 0.11 \pm 0.08 $  \\[.2ex]
    J1335+08 & $ 1.07 \pm 0.25 $ & $ 0.78 \pm 0.25 $  & $ 1.36 \pm 0.29 $ & $ 0.91 \pm 0.31 $ & $ 1.04 \pm 0.36 $  \\[.2ex]
    J1457+22 & -                 & -                  & $ 0.85 \pm 0.19 $ & $ 0.62 \pm 0.21 $ & $ 0.73 \pm 0.21 $  \\[.2ex]
    J1509+37 & $ 0.53 \pm 0.08 $ &  $ 0.35 \pm 0.06 $ & -                 & $ 0.34 \pm 0.07 $ & $ 0.32 \pm 0.07 $  \\[.2ex]
    J0851+21 & $ 0.76 \pm 0.15 $ & -                  & $ 0.93 \pm 0.2 $  & $ 0.45 \pm 0.1 $  & $ 0.48 \pm 0.1 $   \\[.2ex]
    J0213+00 & $ 0.45 \pm 0.11 $ & $ 0.68 \pm 0.21 $  & $ 0.51 \pm 0.14 $ & $ 0.22 \pm 0.12 $ & $ 0.15 \pm 0.13 $  \\[.2ex]
    J1226+04 & -                 &   -                & $ 0.55 \pm 0.15 $ & $ 0.24 \pm 0.1 $  & $ 0.28 \pm 0.15 $  \\[.2ex]
    J0240-08 & $ 0.68 \pm 0.37 $ & $ 0.61 \pm 0.3 $   & $ 0.69 \pm 0.39 $ & $ 0.34 \pm 0.21 $ & $ 0.41 \pm 0.2 $   \\[.2ex]
    J1735+57 & $ 0.70 \pm 0.17 $ & $ 0.47 \pm 0.13 $  & $ 0.71 \pm 0.27 $ & $ 0.33 \pm 0.14 $ & $ 0.49 \pm 0.37 $  \\[.2ex]
    J2302+00 & $ 0.15 \pm 0.09 $ & $ 0.09 \pm 0.07 $  & -                 & -                 & $ 0.31 \pm 0.29 $  \\[.2ex]
    J1311+00 & $ 0.44 \pm 0.12 $ & $ 0.22 \pm 0.11 $  & $ 0.35 \pm 0.17 $ & $ 0.26 \pm 0.15 $ & $ 0.28 \pm 0.22 $  \\[.2ex]
    J1200+27 & $ 0.4 \pm 0.2 $   & $ 0.13 \pm 0.15 $  & $ 0.17 \pm 0.13 $ & -                 & $ 0.19 \pm 0.2 $   \\[.2ex]
    J0808+17 & $ 0.14 \pm 0.4 $  & $ 0.05 \pm 0.33 $  & -                 & -                 & -   \\[1ex]
    J0815+58 & -                 & -                  & -                 & -                 & - \\[1ex]
    J0303-07 & -                 & -                  & $ 0.93 \pm 0.48 $ & $ 0.52 \pm 0.39 $ & $ 0.92 \pm 0.45 $  \\[1ex]
    J1219+15 & -                 & -                  & -                 & -                 & -  \\[1ex]
 \enddata
\end{deluxetable*}

\subsection{Low-Ionization Interstellar Absorption Lines}
For an independent measurement on covering fractions and column densities, we use low-ionization state (LIS) metal lines with ionization potentials less than 13.6 eV to infer covering fractions and column densities for our sample. Our \textit{HST} spectra show numerous nebular and stellar emission lines; we restrict our current analysis to Si II and O I because of their low ionization potentials. Si II is the dominant ion of silicon in the ISM and is commonly used to trace neutral gas geometry and kinematics (e.g. \citealt{Heckman2011, RivThorsen2015,Chisholm2018, Gazagnes2018}). We also observe multiple Si II lines with different strengths which provide robust constraints on line saturation. Specifically, we use Si II and O I to calculate covering fractions, assess optical depth and to constrain neutral gas column densities. 

In general, the residual intensity within an absorption line is sensitive to the column density of absorbing material as well as the covering fraction of a background source of light. Comparing multiple lines with different oscillator strengths $f\lambda$ can lift this degeneracy \citep{savage}. We consider two scenarios: an optically thin shell with uniform coverage and optically thick clouds interspersed with optically thin channels. If the gas is optically thin, the depth of an absorption line depends on the species' oscillator strength, and equivalent widths grow $\propto N f \lambda^2$ on the linear part of the curve of growth. In particular the ratio between equivalent widths of two optically thin lines is given by $EW_i/EW_j = f_i \lambda_i^2 / f_j \lambda_j^2$, assuming complete coverage of the background source.

Another scenario is that the background source is only partially covered by optically thick clouds and radiation escapes through optically thin channels (e.g. the ``picket-fence'' model; \citealt{Heckman2011}). In this case, the residual intensity depends primarily on the covering fraction. Equivalent width ratios grow independently of $f_i\lambda_i^2$, and line profiles appear identical across different transitions. Equivalent width ratios are therefore a powerful tool in discriminating between these optically thin and thick conditions. 

We study five LIS lines found in most of our COS spectra: O I $\lambda1302$, Si II $\lambda1190$, $\lambda1193$, $\lambda1260$, and $\lambda1304$. For each transition, we calculate equivalent widths by integrating 
\begin{equation}
EW\equiv \int \Big(1 - \frac{F}{F_{0}}\Big)  d\lambda
\label{ew}
\end{equation}
over the wavelength range where the line falls below the continuum level. $F$ and $F_0$ are the absorption line and continuum fluxes respectively. Equivalent widths are shown in Table \ref{ews}. Column densities of each species are evaluated by 
\begin{equation}
N = \frac{m_e c}{\pi e^2 f \lambda} \int \ln \frac{F}{F_{0}} d\lambda
\label{optdepth}
\end{equation}
over the same wavelength range. Error bars are estimated for equivalent widths and column densities through Monte Carlo simulations where the same analysis is repeated 1000 times after perturbing each flux point by a random normal deviate drawn from the spectral noise distribution and allowing the continuum to fluctuate randomly by 10\%. We take the standard deviation of the distribution of derived quantities as the uncertainty. Table \ref{all_columns} includes all LIS column density measurements. 

It was necessary to estimate J1448's continuum in the region of Si II $\lambda1190,1193$ accounting for overlap with Milky Way Ly$\alpha$ absorption. In practice this involved fitting a Voigt profile to the Milky Way absorption region and stitching that result to the best-fit model already established from J1448's Ly$\alpha$ MCMC fit. We calculate $N$ and $EW$ as before for J1448, still allowing for 10\% continuum uncertainty. 

\begin{figure}[tb]
    \centering
    \includegraphics[width=\linewidth]{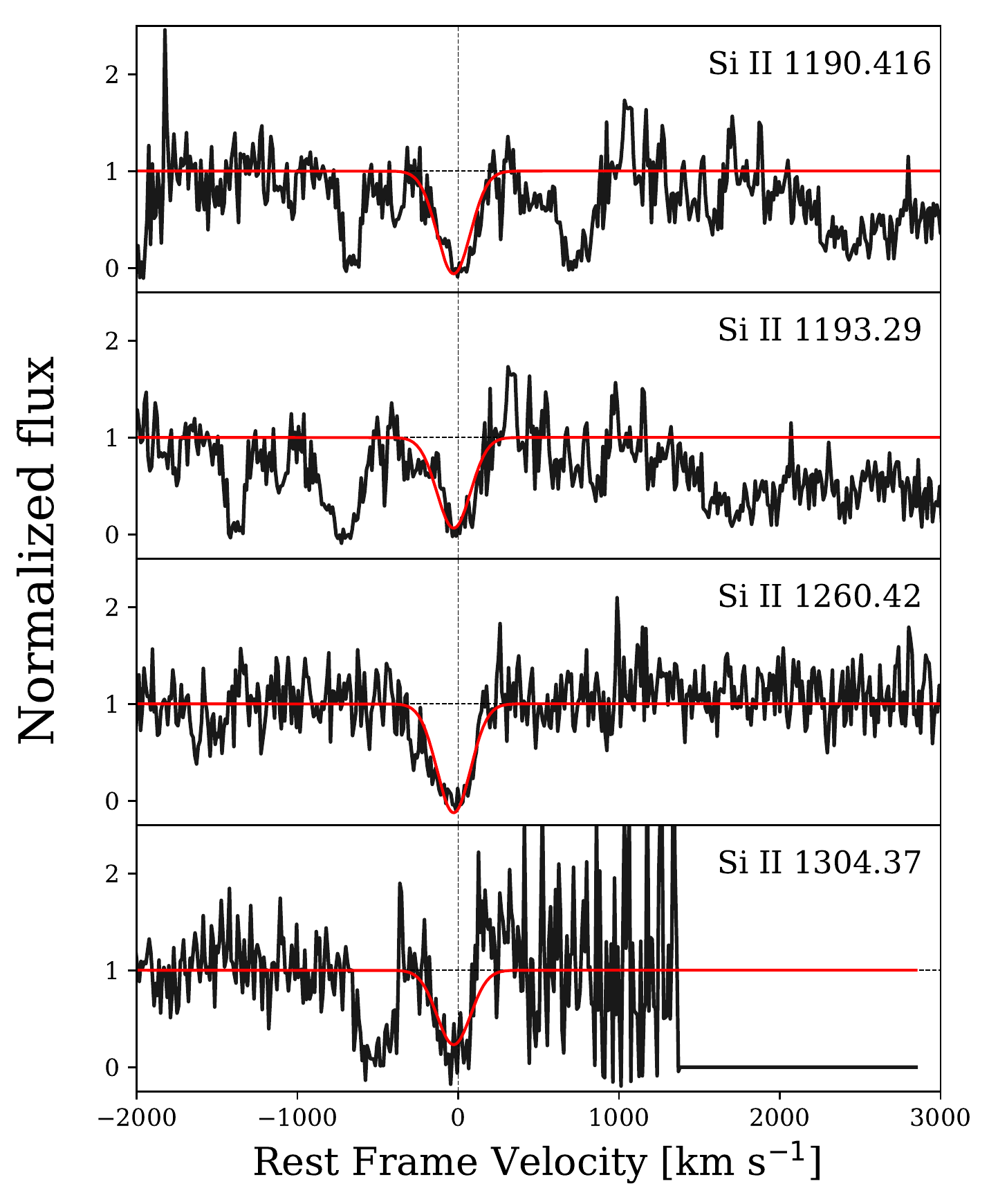}
    \caption{Normalized Si II low-ionization absorption line profiles for J1335 plotted at each line's rest frame velocity. Gaussian fits to each line are shown in red. The fit maintains a constant velocity offset and width for each line but allows the individual depths to vary. Vertical solid black lines indicate a rest-frame velocity of 0 km/s while horizontal black lines show a normalized flux of unity.}
    \label{LIS_abs}
\end{figure}
\subsection{Covering Fractions} 

The equivalent width ratios in our sample are consistent with saturation. This suggests that the non-zero fluxes at absorption line centers are due to radiation escaping through optically thin channels. These results agree with the gas properties of Lyman break analogs as well as confirmed LyC-emitters \citep{Heckman2011,Gazagnes2018,Chisholm2018}. We therefore consider the \cite{Heckman2011} ``picket-fence'' model where optically thick clouds cover a fraction \fcov\ of the ionizing source on the sky. For saturated transitions and low dust content, the covering fraction is directly related to the residual intensity at the central wavelength $\lambda_c$ of each absorption profile by

\begin{equation}
\mathcov(\lambda_c) = 1 - \frac{F(\lambda_c)}{F_0(\lambda_c)}
\label{cf_0}
\end{equation}

We infer Si II covering fractions following \cite{Heckman2011} by simultaneously fitting a single Gaussian absorption profile to all available Si II species in the \textit{HST} COS spectra. We keep Gaussian widths and velocity offsets the same for each transition but allow the depths to vary independently. We estimate \fcov\ by applying Equation \ref{cf_0} to the Gaussian fits and averaging over all available Si II lines. This technique is shown for J1335 in Figure \ref{LIS_abs}. Average values are within the uncertainty of individual measurements, and error bars are calculated using the Monte Carlo techniques discussed previously. All fits were calculated with Levenberg-Marquardt $\chi^2$-minimization implemented in the Python package \texttt{lmfit} \citep{lmfit}. In some cases the observed Si II covering fraction may be marginally $>1$. We attribute this to either observational noise or uncertainty introduced during continuum normalization. Therefore, we interpret all covering fractions above unity as at most equal to one. 

We calculate $3\sigma$ upper limits on column densities and covering fractions for J0808, J0815 and J1219, for which no LIS lines are detected in absorption. To estimate the upper limits on \fcov, we calculate the average flux error per pixel in the normalized spectrum near each Si II transition and adopt this value as the residual intensity. For column density upper limits in Table \ref{all_columns}, we approximate the integral in Equation \ref{optdepth} by using three times the average error per pixel and assuming a line width equal to the average FWHM of the sample: 225 km/s and 187 km/s for Si II and O I respectively. 

\section{Neutral Gas in Extreme Green Peas}\label{results}
\subsection{HI Column Densities}
There are significant neutral gas column densities in GPs with Ly$\alpha$ absorption. Among the eight modeled GPs, all have $\log N_{\mathrm{HI}}/$cm$^{-2} > 17$, the optically thin limit for LyC photons. Seven GPs have column densities greater than $10^{20}$ cm$^{-2}$. Furthermore, we find additional evidence for optically thick gas in the MCMC models; Doppler parameters have no impact on the goodness of fit suggesting saturated, damped absorption. As a result, we find evidence for large neutral gas reservoirs in all modeled GPs, $75\%$ of which show significant Ly$\alpha$ escape as well. 

HI column densities are larger on average in GPs with no Ly$\alpha$ emission. We calculate $\log N_{\mathrm{HI}}/\mbox{cm}^{-2} \gtrsim 21.2$ for J1448 and J1335, two sources with Ly$\alpha$ only in absorption. These GPs also have the highest LIS equivalent widths seen in the sample, and Si II covering fractions near unity. However, large HI column densities are not restricted to $\mathcov\sim1$. We estimate an HI column density for J1608 of $\log N_{\mathrm{HI}}/\mbox{cm}^{-2} = 21.4$ and a Si II covering fraction of $\mathcov=0.42$. Thus, GPs with HI column densities $\log N_{\mathrm{HI}}/\mbox{cm}^{-2}>21$ like J1608 are not necessarily ionization-bounded, as would otherwise be expected from $\log N_{\mathrm{HI}}/\mbox{cm}^{-2}>17$. 

Neutral gas absorbs LyC and scatters Ly$\alpha$. As a result, we expect to find greater Ly$\alpha$ escape fractions at lower HI column densities. To assess this relation with our entire sample, we estimate \NHI\ using Si II and OI column densities to supplement our Ly$\alpha$ results. Si II can be used to estimate a rough HI column density with each GPs' oxygen abundance by assuming all Si is in the form of Si II and on average $\log$(Si/O)=1.59, characteristic of extragalactic HII regions with similar metallicities to the GPs \citep{Garnett1995}. We do this for each transition and average the results for every target with Si II detected in absorption. Individual \NHI\ estimates from various Si II transitions are typically within $1\sigma$ of each other for a given GP. To estimate \NHI\ from O I we make a similar assumption but need to use only the measured $12+\log$(O/H). \NHI\ derived from both Si II and O I are shown in Table \ref{derived_n}. 

\begin{deluxetable}{lcc}
    \caption{Derived HI Column Densities from LIS lines}
    \label{derived_n}
    \tablehead{GP  &$\log$\NHI(Si II)/cm$^{-2}$ &$\log$\NHI(O I)/cm$^{-2}$  } 
    \startdata 
    J1448-01 & $19.89\pm0.26$ & $19.16\pm0.25$ \\[.2ex]
    J1608+35 & $19.64\pm0.25$ & $18.22\pm0.31$ \\[.2ex]
    J1335+08 & $20.62\pm0.34$ & $19.53\pm0.60$ \\[.2ex]
    J1457+22 & $20.02\pm0.25$ & $18.83\pm0.58$ \\[.2ex]
    J1509+37 & $20.06\pm0.22$ & $18.91\pm0.15$ \\[.2ex]
    J0851+21 & $20.17\pm0.31$ & $19.10\pm0.42$ \\[.2ex]
    J0213+00 & $19.63\pm0.27$ & \---           \\[.2ex]
    J1226+04 & $19.59\pm0.22$ & $18.65\pm0.35$ \\[.2ex]
    J0240-08 & $19.62\pm0.24$ & \---           \\[.2ex]
    J1735+57 & $19.51\pm0.24$ & $18.55\pm0.17$ \\[.2ex]
    J2302+00 & $19.78\pm0.14$ & \---   \\[.2ex]
    J1311+00 & $19.37\pm0.30$ & \---   \\[.2ex]
    J1200+27 & $18.98\pm0.19$ & \---   \\[.2ex]
    J0808+17 & \--- & \---   \\[1ex]
    J0815+58 & \--- & \---    \\[1ex]
    J0303-07 & $20.02\pm0.13$ &  $19.21\pm0.46$   \\[1ex]
    J1219+15 & \--- & \---   \\[1ex]
 \enddata
\end{deluxetable}

We find that on average, $f_{esc}^{Ly\alpha}$ tends to decrease with greater \NHI\ for each method. Figure \ref{fesc_HI} compares $f_{esc}^{Ly\alpha}$ with HI column densities derived from Ly$\alpha$ absorption fitting, Si II, and OI. We find that $f_{esc}^{Ly\alpha}$ decreases steeply from $\log N_{\mathrm{HI}}/$cm$^{-2} =19.5-20.5$. J1608 is a notable outlier with a high HI column density of $\log N_{\mathrm{HI}}/$cm$^{-2}=21.39$ for its Ly$\alpha$ escape fraction of only 18\%. Otherwise, GPs with $\log N_{\mathrm{HI}}/$cm$^{-2}>20$ generally have $f_{esc}^{Ly\alpha}<5\%$. Discrepancies between HI column density measurements from different techniques are discussed in the following section.
     
\subsection{Low-ionization Column Densities and Equivalent Widths}\label{LISs}

Low-ionization column densities and equivalent width ratios are consistent with optically thick gas in our sample of GPs. In particular, we find that Si II EW ratios generally disagree with optically thin predictions regardless of whether or not Ly$\alpha$ is detected in absorption. The most robust evidence for saturation comes from absorption line pairs where $f\lambda$ ratios differ by at least a factor of two \citep{savage}. Si II $\lambda1193/\lambda1304$ and $\lambda1260/\lambda1304$ have sufficiently different $f_i\lambda_i^2/f_j\lambda_j^2>5$. In GPs with both transitions detected, these EW ratios are near unity, and deviate from optically thin predictions by $\gtrsim3\sigma$. Thus, Si II absorption lines are saturated and any residual intensity at the line's core is likely due to low covering fraction. 

\begin{figure}[tb]
    \centering
    \includegraphics[width=\linewidth]{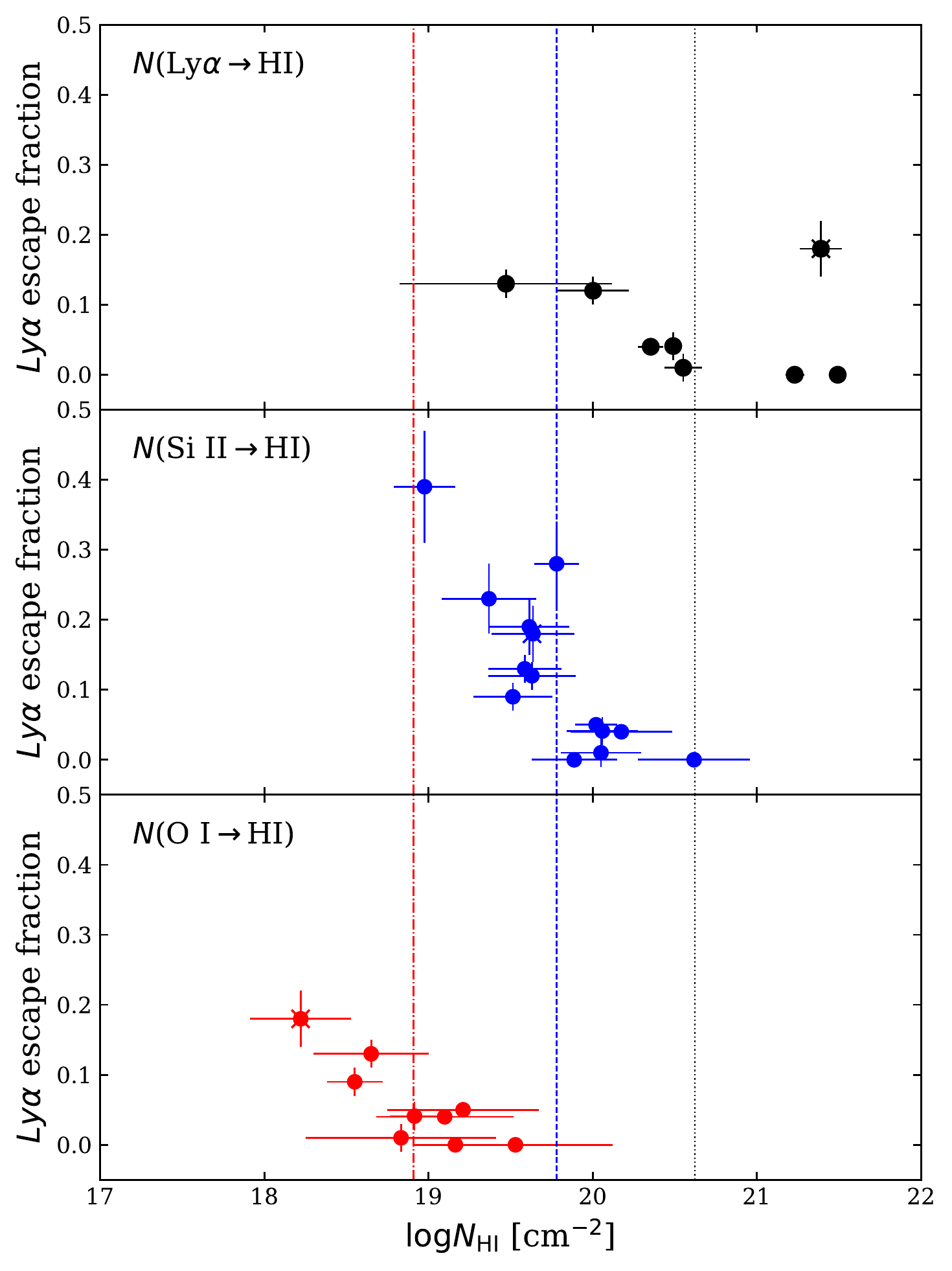}
    \caption{Ly$\alpha$ escape fraction from \cite{Jaskot2017} vs. HI column density estimated with three different techniques for the same sample of extreme GPs. Results of MCMC Voigt fits to Ly$\alpha$ absorption wings are shown in the top panel as black circles. The vertical dotted black line indicates the average N(Ly$\alpha\rightarrow$HI) value. Blue (red) circles correspond to Si II (O I) estimates of \NHI\ in the middle (bottom) panel for GPs with detected LIS absorption lines. The vertical dashed blue line and dot-dashed red line indicate average values for N(Si II$\rightarrow$HI) and N(O I$\rightarrow$HI) respectively.}. 
    \label{fesc_HI}
\end{figure}

Ly$\alpha$ and LIS absorption lines indicate potentially large \NHI\ along the line of sight, however, derived \NHI\ values for a given GP can vary significantly between different tracers. Figure \ref{fesc_HI} shows \fesc\ vs. derived \NHI\ from Ly$\alpha$, O I and Si II absorption. Ly$\alpha$ and Si II tend to agree on large $\log$\NHI/cm$^{-2}\sim 19-21$ within 1$\sigma$ errors for a given GP, however, the average Si II- (O I)- derived \NHI\ is offset from Ly$\alpha$ results by $\sim1$ dex (2 dex). On one hand, observing Ly$\alpha$ absorption may be biased towards the highest \NHI\ systems. Furthermore, strong Ly$\alpha$ emission could fill in Ly$\alpha$ absorption in GPs with low \fcov\ and low HI optical depth. Thus, the average \NHI(Ly$\alpha$) may be greater than that of Si II and O I by virtue of missing GPs with low \NHI. Alternatively, Si II and O I may underestimate column densities due to depletion onto dust or radiative infilling, a process by which saturated absorption line cores are filled in by LIS photons emitted into the line of sight. Covering fractions below unity may increase the observed flux inside absorption lines, decreasing the column density calculated with Equation \ref{optdepth}. For these reasons, and due to uncertainty on Si/O abundance ratios as well as the presence of other Si species in the COS spectra, Si II-derived \NHI\ should be considered as a lower limit. 

O I column densities are $>10^{18}$ cm$^{-2}$ when detected but consistently below \NHI\ estimates from other absorption lines. Si II $\lambda1260$ and O I $\lambda1302$ absorption line strengths should scale similarly due to comparable transition probabilities of the Si II $\lambda1264$* and O I $\lambda1304$* lines that relieve the effects of radiative infilling. Correcting column density measurements by each line's infilling probability does not resolve the differences between Si- and O-derived \NHI; lines of similar infilling probability do not agree within $1\sigma$ errors. Alternatively, dust depletion may play an important role even in extremely ionized conditions \citep{Howk1999}. \cite{Jenkins2009} finds large oxygen depletion losses compared to Si in the local Milky Way, which could explain the differences between Si II and O I derived \NHI\ in the GPs. On the other hand, we note that the line saturation of O I is unconstrained by only one observed transition. Thus, O I $\lambda1302$ may not accurately trace the O I column density. \NHI\ estimates using either O or Si may not agree, and the use of only one diagnostic may poorly constrain the neutral gas content.

\subsection{Si II Covering Fractions}
     
\begin{figure*}[tb!]
    \begin{minipage}{.48\textwidth}
    \centering
    \includegraphics[width=\linewidth]{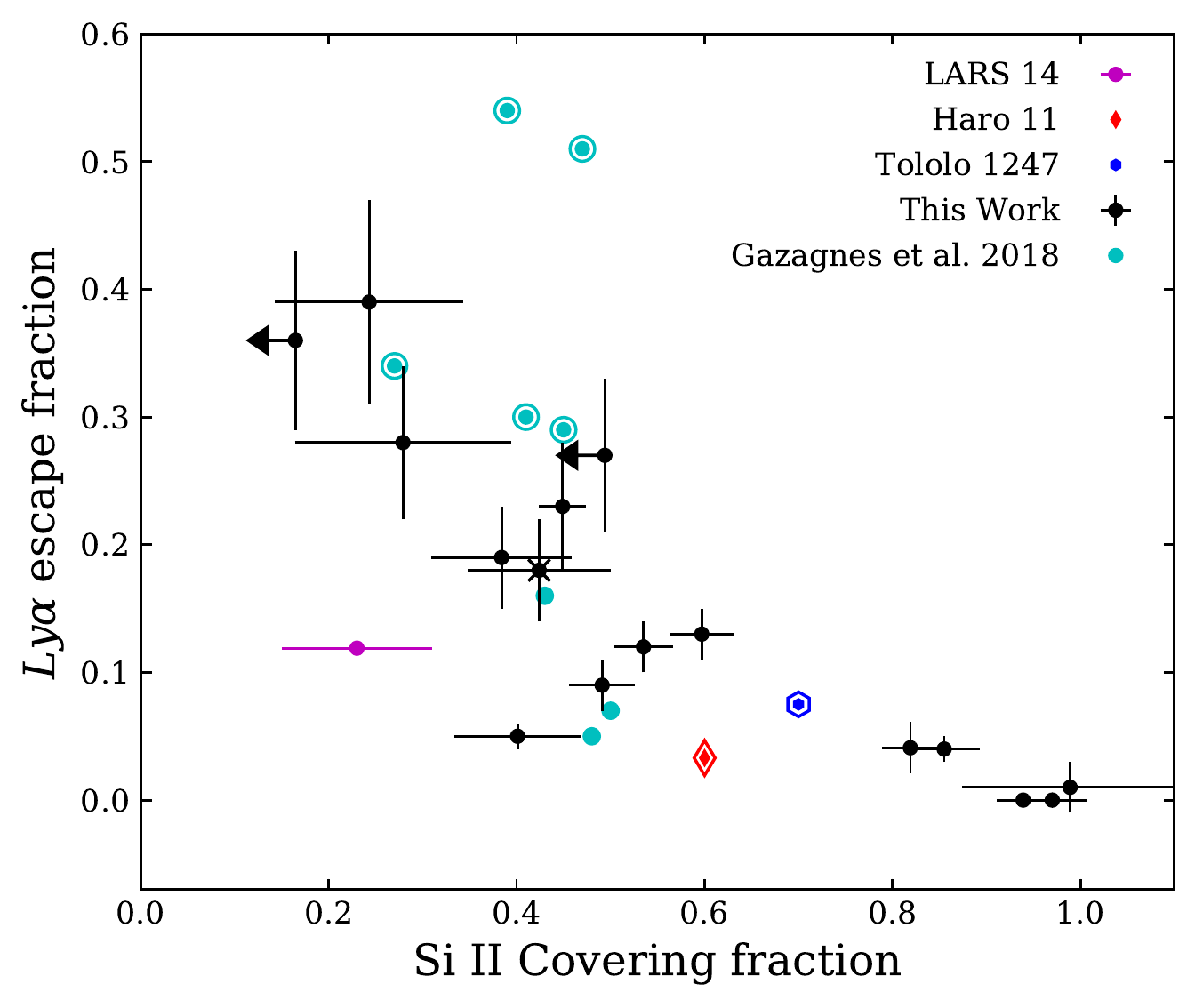}
    \end{minipage}
    \begin{minipage}{.48\textwidth}
    \centering
    \includegraphics[width=\linewidth]{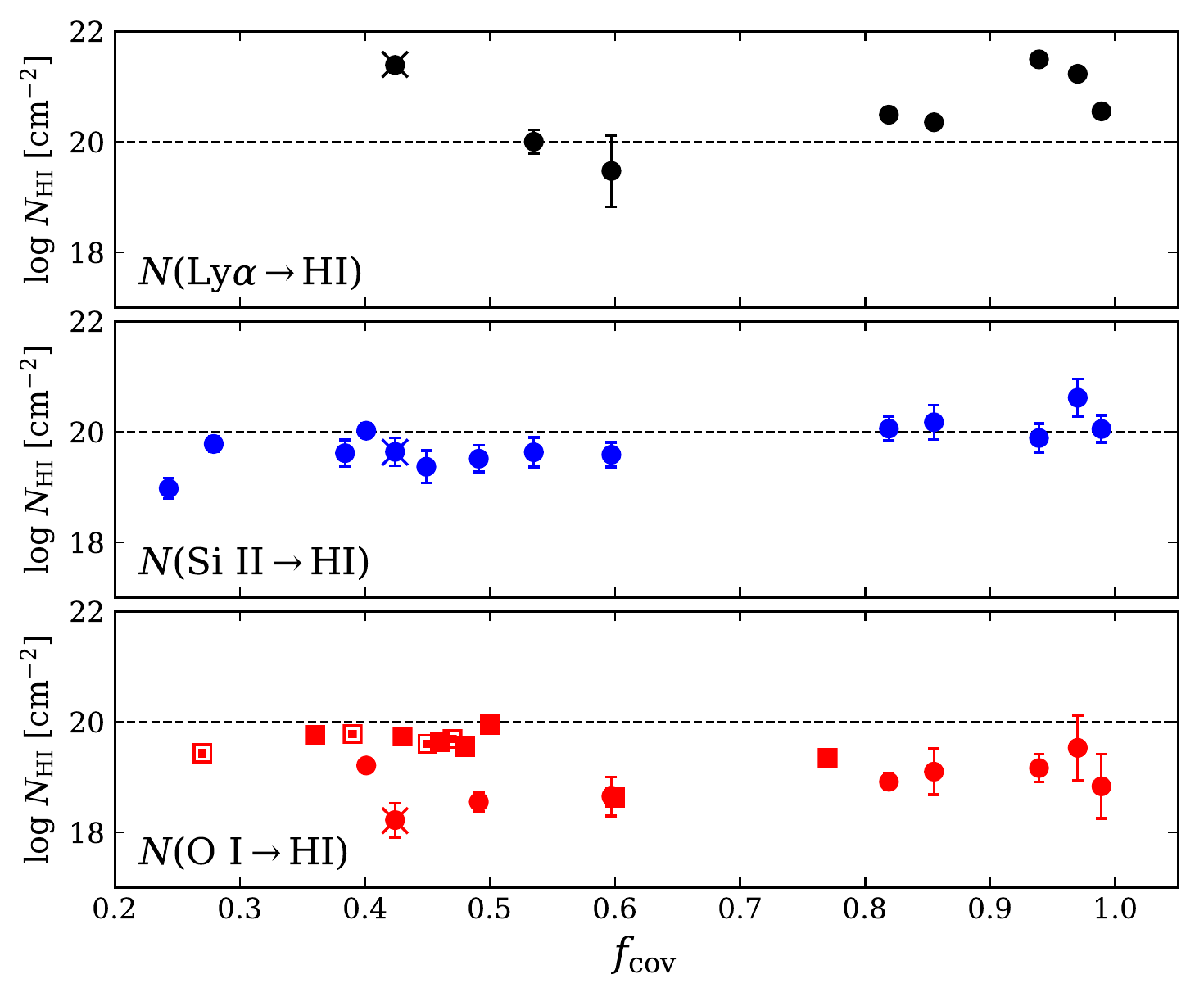}
    \end{minipage}
    \caption{(\textit{Left}) Ly$\alpha$ escape fraction compared to Si II covering fraction. Our data are shown as black circles, with upper limits on \fcov\ for J0808 and J0815 represented by arrows. Cyan circles are measurements from \cite{Gazagnes2018} for a sample of GPs. LARS 14 is shown with a magenta square. Haro 11 and Tololo 1247 are shown in red and blue respectively. Data points surrounded by open markers are confirmed LCEs. J1608 is marked with a $\times$ symbol. (\textit{Right}) HI column density vs. Si II covering fraction. Separate panels are shown for each technique used to estimate \NHI. Data from this work are shown as circles. HI column densities estimated with MCMC fits to Ly$\alpha$ absorption wings are shown in black. Blue and red indicate Si II- and O I- derived HI column densities respectively. \cite{Gazagnes2018} GPs are shown as squares, with open markers indicating LCEs. The dotted horizontal line shows $\log N_{\mathrm{HI}}=20$ on each plot for comparison. }
    \label{covs}
\end{figure*}

\cite{RivThorsen2015} first noticed an anti-correlation between covering fraction and Ly$\alpha$ escape fraction in the Lyman Alpha Reference Sample (LARS, \citealt{Ostlin2014}), a sample of star-forming galaxies selected for strong H$\alpha$ and to span a range of UV luminosities. However, those authors find consistently large covering fractions $\gtrsim0.8$ for all LARS targets except LARS 14, which is a GP. To check for consistency, we download and analyze the COS spectra for LARS 14 from Program 11727 (PI: Heckman). We measure its Si II covering fraction to be $0.23\pm0.08$, marginally lower than the maximum LIS covering fraction of $0.40\pm0.05$ presented in \cite{RivThorsen2015}. The discrepancy between these numbers is related to differences in methodology: \cite{RivThorsen2015} calculate \fcov$(v)$ by averaging Si II, O I and C II absorption profiles, whereas this work only considers Si II transitions. In either case, LARS 14 has a low \fcov\ for its intermediate \fesc$=0.119$.  

As shown in the left panel of Figure \ref{covs}, the GPs' Si II covering fractions range from $0.1-1$, and we find that \fcov\ anti-correlates with $f_{esc}^{Ly\alpha}$. For reference we also show measurements from \cite{Gazagnes2018} who find HI and Si II covering fractions below unity in LCE GPs. In our sample of extreme GPs, the $f_{esc}^{Ly\alpha}-\mathcov$ relation is the strongest anti-correlation found between $f_{esc}^{Ly\alpha}$ and other measured properties such as UV/optical emission line ratios, 12+$\log$(O/H), EW(Ly$\alpha$), EW(H$\alpha$), and offset velocities of low- and high-ionization metal absorption lines. ISM porosity plays a critical role in regulating Ly$\alpha$ photon escape from GPs. 

The observed relationship between high $f_{esc}^{Ly\alpha}$ and low \fcov\ is likely affected by the GPs' orientation along the line of sight. Simulations find that ionizing photon escape is highly anisotropic, and that larger escape fractions correspond to larger opening angles (e.g. \citealt{Paardekooper2015}). In that case, strong Ly$\alpha$ emission may correlate with low covering fraction if more optically thin channels fall in the COS beam. Indeed we recover this trend in the left panel of Fig. \ref{covs}. 

We compare \fcov\ with \NHI\ from Ly$\alpha$, Si II and OI in the right panel of Fig. \ref{covs}. We do not find evidence for a strong trend between \NHI\ and Si II covering fraction. As shown in Fig. \ref{covs}, GPs with column densities from Ly$\alpha$ above $10^{20}$ cm$^{-2}$ may preferentially appear at covering fractions closer to unity. However, we note that J1608 has one of the highest \NHI\ yet $\mathcov\sim0.4$, indicating that not all high column density targets need be completely covered. Si II covering fractions change little with HI column density, suggesting that large neutral gas densities may not be distributed uniformly.

\section{Discussion}\label{discussion}
\subsection{Neutral Gas Geometry}

Two idealized scenarios limit Ly$\alpha$ escape from dense, nebular regions. The ``picket-fence'' model \citep{Heckman2011} describes an ionizing source partially covered by a distribution of optically thick gas clouds. Ly$\alpha$ photons may still resonantly scatter through optically thick regions but the emission profile can be dominated by direct escape \citep{Dubal2014,Verhamme2015}. Metal absorption lines will be saturated, and Equation \ref{cf_0} describes the residual intensity in the core of the line. Another possible scenario is the ``density-bounded'' or ``uniform shell'' model in which the ionizing source is covered completely ($\mathcov=1$) and the ISM must be optically thin for significant escape.  

As shown in Sec. \ref{LISs}, the gas densities and line saturation observed in our sample strongly support the picket-fence model of Ly$\alpha$ escape. Low-ionization line equivalent width ratios disagree with optically thin predictions for all available transitions in every target. Voigt profile fits are all insensitive to the Doppler parameter $b$ which is consistent with saturation on the curve of growth. We therefore rule out the uniform shell model in our GPs that show metal absorption lines. Some GPs have low \fcov, suggesting that they may be LCEs despite evidence for optically thick gas. 

Within the picket-fence model, \cite{RivThorsen2015} put forth two scenarios. A clumpy ISM may consist of clouds moving at a single velocity driven uniformly by stellar feedback. In this case, ionizing photons escape according to the standard picket-fence model through optically thin channels. An alternate scenario pictures multiple clouds at different velocities, each only partially covering the ionizing source yet together covering it completely. This may arise if Rayleigh-Taylor instabilities occur in outflows, causing fragmentation at multiple velocities (e.g. \citealt{Tenorio1999}). Ly$\alpha$ may scatter to velocities with $\mathcov(v)$ less than one and escape out of resonance, in which case the Ly$\alpha$ profile may have enhanced emission away from systemic velocity (e.g. \citealt{Dubal2014}). Observing a low-ionization covering fraction less than unity is still possible because at any given velocity the clouds do not cover the entire source. 

Low systemic Ly$\alpha$ emission in the COS spectra suggests that low-column density regions must have enough HI gas to scatter Ly$\alpha$. Moreover, LIS absorption profiles are kinematically aligned across different species, and offset from line center by $<40$ km/s. The standard deviation of cloud velocities are consistently $\leq 100$ km/s. Thus, clouds distributed over a large velocity range are unlikely, and we favor a picket-fence scenario with low-column density channels and velocities $\leq 100$ km/s. 

Evidence for high HI column densities and covering fractions below unity suggests that Ly$\alpha$ photons scatter and escape through optically thin channels in the ISM. We consistently find low Si II covering fractions in targets with greater $f_{esc}^{Ly\alpha}$. Furthermore the $\mathcov-f_{esc}^{Ly\alpha}$ correlation is high, suggesting that low \fcov\ may be the most important criterion for observing Ly$\alpha$ escape in our sample. This result should be expected if a picket-fence model with low-column density channels describes the GPs' geometry. If this is the case then Ly$\alpha$ escape is likely anisotropic (e.g. \citealt{Dove2001,Gnedin2008,Zastrow2011,CenKimm2015}). We speculate that J1335 and J1448, which have no Ly$\alpha$ emission along the line of sight, could be Ly$\alpha$ emitters if viewed from a different angle. 

Prior studies have proposed mechanical feedback as a means of producing low covering fractions. In particular, supernovae (SNe) may play an important role in enhancing LyC escape.  \cite{ClarkeOey2002} suggest that quasi-adiabatic SN-driven bubbles are capable of clearing out HI gas and boosting escape fractions in regions where the SFR exceeds a critical value. High resolution hydrodynamic simulations of dwarf galaxies similarly show that ionizing photons preferentially escape through low column density channels cleared out by SN-driven outflows \citep{WiseCen2009,Trebitsch2017}.

\begin{figure}[tb!]
    \centering
    \includegraphics[width=\linewidth]{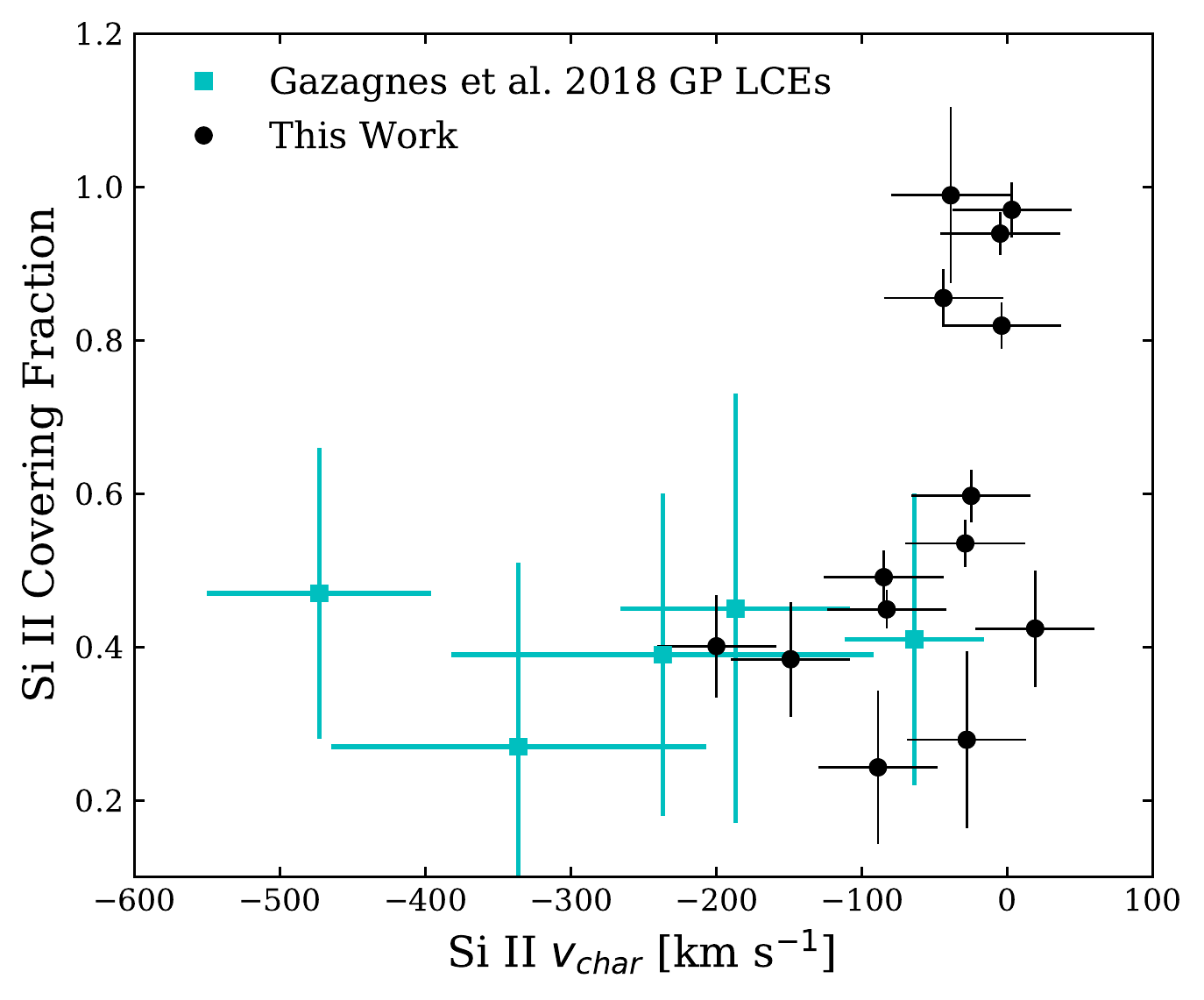}
    \caption{GP Si II covering fraction vs. $v_{char}$, the velocity weighted by absorption line depth and averaged over all detected transitions. Our data are shown as black circles. Cyan points are GP LCEs taken from \cite{Chisholm2017} and \cite{Gazagnes2018}.}
    \label{fc_vcen}
\end{figure}

However, recent observational studies suggest that other feedback mechanisms may be at work in GPs. \textit{HST} COS observations of GPs find little or no correlation between outflow velocities and Ly$\alpha$ escape. In a sample of 10 targets \cite{Henry2015} find no correlation between Ly$\alpha$ escape fractions and outflow velocities traced by Si II and C II low-ionization transitions. Furthermore, \cite{Chisholm2017} find no extreme velocities in a sample of LCEs. \cite{Jaskot2017} demonstrate that while superwind velocities may correlate with low optical depths in some cases, highly ionized GPs showing strong, narrow Ly$\alpha$ emission and weak low-ionization absorption also have the lowest wind velocities. These extreme GPs require feedback mechanisms beyond SN-driven outflows, such as radiative feedback or mechanical feedback from a prior generation of stars \citep{Micheva2017}. 

Figure \ref{fc_vcen} compares $v_{char}$, the Si II velocity weighted by absorption line depth and averaged over all detected transitions, with \fcov\ in extreme GPs and GP LCEs. High covering fractions are exclusively found at low $v_{char}$ in our sample, and low covering fractions appear at negative velocities in some GPs, suggesting that mechanical feedback in extreme GPs clears out gas when present. However, we note that other samples may not show this trend. For example, \cite{Heckman2011} find that velocity correlates with low column density in Lyman Break Analogs (LBAs), evidence for starburst-driven outflows in compact regions.  

A different scenario may produce low covering fractions at low velocity in extreme GPs. For example, \cite{Oey2017} find suppressed superwinds in the super star-cluster Mrk 71-A, a GP analog and LCE candidate likely dominated by radiative feedback. On the other hand, that system also shows a two-stage starburst with an older and younger component of ages $<3-5$ Myr and $\lesssim1$ Myr respectively \citep{Micheva2017}. LyC photons can therefore escape through holes cleared out during prior epochs of star-formation. In either case, mechanical feedback from the starburst itself may not be critical to the escape of Ly$\alpha$ and, potentially, LyC radiation from extreme GPs. The obscuring gas may be independent of the GPs' starburst regions and distant from the UV sources along the line of sight.

High-resolution simulations find that SFRs in dwarf galaxies naturally vary on short time scales, which can lead to strong fluctuations in ionizing photon escape fractions (e.g. \cite{WiseCen2009,Hopkins2014,CenKimm2014,Paardekooper2015,Trebitsch2017}). Consequently, LyC escape fractions are transient and highly anisotropic (e.g. \citealt{Ma2015}). J1608, a strong LCE candidate, likely harbors an extremely young stellar population with age $\lesssim3-5$ Myr (e.g. \citealt{Jaskot2017}), driving its high $f_{esc}^{Ly\alpha}$ and $O_{32}$. J1608's high ratio of SFR/$M_{HI}$ is also consistent with a temporary boosting of H$\alpha$ emission. Furthermore, \cite{Jaskot2017} find a lack of evidence for outflows in J1608, consistent with the existence of suppressed superwinds. This suggests that SNe have yet to remove sources of LyC photons, which may manage to escape this environment. Thus, mechanical feedback may not have much influence on the surrounding gas $1-5$ Myr after the starburst, before massive O- and B-type stars evolve off the main-sequence.

\begin{figure*}[tb!]
    \centering
    \includegraphics[width=\linewidth]{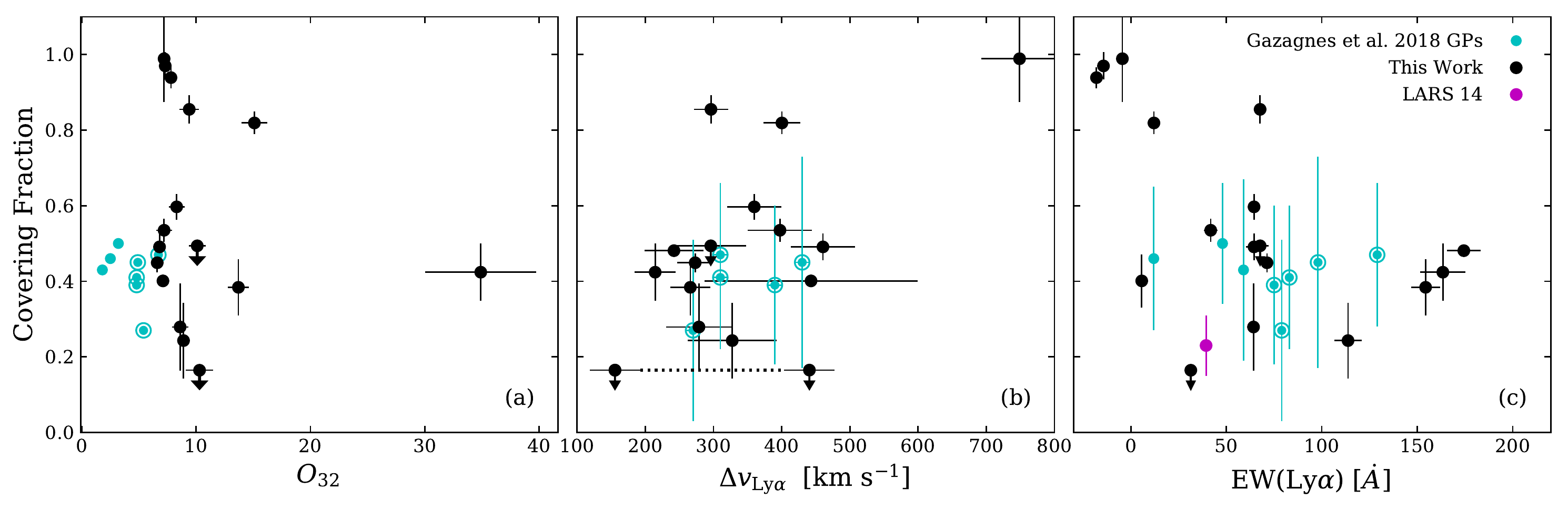}
    \caption{Si II covering fraction compared to various tracers of Ly$\alpha$ escape fractions. Our data are shown as black circles. Covering fraction upper limits for J0808 and J0815 are indicated with arrows. LARS 14 is shown in magenta and \cite{Gazagnes2018} GPs are shown in cyan. Circles indicate GPs, and rings denote LCEs. \textit{(a)} Si II covering fraction vs. $O_{32}=[\mbox{O}\mathrm{III}]/[\mbox{O}\mathrm{II}]$. \textit{(b)} Si II covering fraction vs. velocity separation between Ly$\alpha$ emission peaks. A dotted black line connects the two values for J0808, a triple-peaked system. \textit{(c)} Si II covering fraction vs. Ly$\alpha$ equivalent width.  
    }
    \label{tracers}
\end{figure*}

\subsection{Does Low \fcov\ Correlate with Predictors of LyC Escape?}

Numerous nebular emission diagnostics have been proposed as indicators of LyC escape and are often used to select LCE candidates. One such tracer is the Ly$\alpha$ escape fraction, which correlates with LyC escape fraction in observations and models (e.g. \citealt{Gronke2015,Verhamme2015,Verhamme2017}). Given that low Si II covering fractions select high $f_{esc}^{Ly\alpha}$, we now investigate whether or not \fcov\ scales with other LCE predictors including $O_{32}$, the velocity separation of Ly$\alpha$ emission peaks ($\Delta v_{\mbox{\scriptsize Ly}\alpha}$) and EW(Ly$\alpha$). 

$O_{32}$, which depends on the hardness of ionizing radiation and ionization parameter, has been suggested as a tracer of LyC escape \citep{JaskotOey2013,NakajimaOuchi2014}. While $O_{32}$ has been found empirically to correlate with LyC escape (e.g. \citealt{Izotov2016c,Izotov2018b}), high $O_{32}$ does not guarantee high LyC escape \citep{Izotov2018,Naidu2018}. Figure \ref{tracers}a compares \fcov\ and $O_{32}$. We do not find a correlation between \fcov\ and $O_{32}$ in extreme GPs. This scatter may be related to an orientation bias, if HI gas is distributed inhomogeneously around the ionizing star forming regions. 

\subsection{The Role of Mechanical Feedback}

In addition to nebular emission diagnostics, the velocity separation of double-peaked Ly$\alpha$ profiles may anti-correlate with LyC and Ly$\alpha$ escape \citep{Verhamme2015,Henry2015,Izotov2018}. As optical depths increase, Ly$\alpha$ photons must undergo more scattering events to Doppler shift out of resonance. Conversely, physical channels through the ISM may facilitate Ly$\alpha$ escape with fewer scattering events. Thus, we might expect that low covering fractions could appear with low $\Delta v_{\mbox{\scriptsize Ly}\alpha}$. We find a possible correlation between these parameters, with large dispersion (Fig. \ref{tracers}b). Covering fraction is sensitive to the distribution of high-column density gas, whereas $\Delta v_{\mbox{\scriptsize Ly}\alpha}$ may preferentially trace lower column density channels and can also be sensitive to the overall 3D distribution of gas, not just gas along the line of sight \citep{Jaskot18}. \cite{Henry2015} argue that a strong anti-correlation between $\Delta v_{\mbox{\scriptsize Ly}\alpha}$ and $f_{esc}^{Ly\alpha}$ in GPs indicates that HI column density is the dominant factor determining Ly$\alpha$ escape. However, our results demonstrate that the relationship between $\Delta v_{\mbox{\scriptsize Ly}\alpha}$, \NHI\ and $f_{esc}^{Ly\alpha}$ is likely complicated by non-unity covering fractions. 

Ly$\alpha$ equivalent widths are useful in selecting LCE candidates at high redshift (e.g. \citealt{Steidedl2018}). Furthermore, we expect to recover a trend between EW(Ly$\alpha$) and covering fraction given that greater $f_{esc}^{Ly\alpha}$ naturally implies higher equivalent width. Indeed, \cite{Verhamme2017} find that EW(Ly$\alpha$) scales with $f_{esc}^{Ly\alpha}$ in GPs and LCEs. In Figure \ref{tracers}c we show that high EW(Ly$\alpha$) tends to favor low covering fraction in extreme GPs, although with large scatter. \cite{Steidedl2018} find a similar trend in stacked LAEs at $z\sim3$, suggesting that EW(Ly$\alpha$) may be a useful indicator of neutral gas geometries favorable for ionizing photon escape at higher $z$.

While we find that high $O_{32}$, low $\Delta v_{\mbox{\scriptsize Ly}\alpha}$ and high EW(Ly$\alpha$) enhance the probability of measuring low covering fractions in GPs, no single diagnostic strongly correlates with \fcov. Our data indicate that a simple spherical model for ionized regions in GPs is unlikely, and predictors for LyC escape will likely need to be multidimensional. 

\section{Summary and Conclusion}\label{conclusion}
We have presented VLA 21cm observations of the GP J1608+35, which has the largest [O III$\lambda5007$]/[O II$\lambda3727$] for SDSS star forming galaxies, and high-resolution UV COS observations of 17 extreme Green Pea (GP) galaxies. Significant Ly$\alpha$ emission is detected in 15 out of 17 targets, with Ly$\alpha$ absorption showing up in eight of the 17 GPs. High inferred Ly$\alpha$ escape fractions ($f_{esc}^{Ly\alpha}$) and large $O_{32}$ make many of these GPs good candidates for escaping Lyman continuum radiation. Their study sheds light on the manner in which Ly$\alpha$ and ionizing radiation escape dense, star-forming regions. The main results of this paper are as follows. 

We do not detect 21cm emission in J1608, and place a $3\sigma$ upper limit on the HI mass of $\log M_{HI}/M_\odot <8.14$. This limit is consistent with the HI content of blue compact dwarfs and HI-selected dwarf galaxies of comparable stellar mass. J1608 has an anomalously high specific star-formation rate for its HI mass, similar to LyC-emitters Tololo 1247-232 and Haro 11. We constrain J1608's HI mass fraction ($f_{HI}\equiv M_{HI}/M_*$) to be $\leq12.56$, falling below predicted values from optical and UV-derived scaling relations. J1608 is likely experiencing a brief period of intense star-formation. A young stellar age could boost H$\alpha$ emission and hence the H$\alpha$-derived SFR, driving the disconnect between J1608's observed $f_{HI}$ and predictions from scaling relations. Like confirmed LyC-emitters, J1608 has an usually high SFR/$M_{HI}$, consistent with a scenario where large ratios of young stars to neutral gas may facilitate stellar feedback and subsequently LyC escape.  
    
We fit the Ly$\alpha$ absorption wings in the GPs' COS spectra and infer HI column densities in the range $\sim10^{19}-10^{21}$ cm$^{-2}$. These values are $2-4$ orders of magnitude above the limiting column density at which gas becomes optically thick to LyC photons. The presence of high column densities is also supported by equivalent width and apparent optical depth analyses of low-ionization absorption lines. Si II absorption lines are saturated, and \NHI\ estimates from O I are systematically lower than Si II-derived  \NHI\ by a factor of 10 or more. 
    
Si II covering fractions $(\mathcov)$, defined as the fraction of optically thick lines of sight in a beam, are as low as $0.2$ in systems with $f_{esc}^{Ly\alpha}>30\%$. We also find a significant anti-correlation between $f_{esc}^{Ly\alpha}$ and covering fraction, consistent with the results of \cite{Chisholm2017} and \cite{Gazagnes2018}. In some GPs, we see both Ly$\alpha$ absorption with HI column densities $>10^{19}$ and strong, narrow Ly$\alpha$ emission, implying lower \NHI. A non-uniform gas covering may explain how these observations appear in the same objects by having optically thin channels through dense regions in the same line of sight. Thus, Ly$\alpha$ escape in the GPs may be anisotropic and detection of Ly$\alpha$ emission could depend on each target's orientation to the line of sight. 
    
GPs with covering fractions close to unity show lower gas outflow velocities in absorption. Low covering fractions appear in GPs over a range of outflow velocities between $-200 < v_{cen} < 0$ km s$^{-1}$ relative to the systemic velocity. We do not find $\mathcov\sim1$ at large negative (blue-shifted) velocity in GPs, suggesting that when mechanical feedback from the ionizing starburst is present, it clears out gas. However, low \fcov\ GPs do not all have strong outflows. While mechanical feedback may operate in some cases, other mechanisms are required to produce low covering fractions in the galaxies with low velocities. Potential candidates include radiative feedback, as suggested by \cite{Jaskot2017} and \cite{Oey2017}, or a two-stage starburst. Ionizing photon escape may be optimized in some GPs at young ages $1-5$ Myr after the starburst, where mechanical feedback may not strongly influence the ISM.
    
To assess whether or not \fcov\ can select LCE candidates, we compare low-ionization covering fractions against tracers of LyC escape used in the literature. We find that high $O_{32}$, high Ly$\alpha$ equivalent widths and low Ly$\alpha$ peak separation favor low covering fractions and may therefore increase the probability of selecting LyC-emitters. However, these relations show significant dispersion and no single diagnostic of LyC-escape strongly correlates with low covering fraction.  

The GPs' gas geometries are complicated and aniso-tropic; a simple density-bounded sphere is ruled out, and orientation may be important in determining which GPs are detected as LyC-emitters. Low Si II covering fractions are the most important criterion for observing high $f_{esc}^{Ly\alpha}$ in highly ionized GPs, and low covering fractions appear at both low and high gas velocities. Low density channels may be optically thin to LyC or low-ionization metal lines like Si II, but not necessarily Ly$\alpha$. Low \fcov\ may play an important role in facilitating LyC escape during the epoch of reionization when neutral gas densities are greater than those observed today. 

\medskip 
\small
We thank the referee for her/his helpful comments, as well as for suggestions that improved the paper. We are grateful to Hansung Gim and Sarah Betti for their helpful advice on VLA data reduction. JM, AEJ and MSO acknowledge support from STScI grant HST-GO-14080. AEJ acknowledges support by NASA through Hubble Fellowship grant HST-HF2-51392. STScI is operated by AURA under NASA contract NAS-5-26555. TD acknowledges support from the Massachusetts Space Grant Consortium. Funding for the Sloan Digital Sky Survey IV has been provided by the Alfred P. Sloan Foundation, the U.S. Department of Energy Office of Science, and the Participating Institutions. SDSS-IV acknowledges support and resources from the Center for High-Performance Computing at the University of Utah. The SDSS web site is www.sdss.org. The National Radio Astronomy Observatory is a facility of the National Science Foundation operated under cooperative agreement by Associated Universities, Inc.

\footnotesize{\bibliography{references.bib}}

\end{document}